\documentclass[preprintnumbers,unsortedaddress,superscriptaddress,notitlepage,square,sort&compress,comma,numbers]{revtex4-1}
\pdfoutput=1
\usepackage{amsfonts}
\usepackage{amsmath}
\usepackage{mathtools}
\usepackage{hyperref}
\usepackage{amssymb}
\usepackage[english]{babel}
\usepackage[utf8]{inputenc}
\usepackage{graphicx}
\usepackage{subcaption}
\usepackage{epsfig}
\usepackage{bm}
\usepackage{verbatim}
\usepackage{float}
\usepackage{wrapfig}
\usepackage{booktabs}
\usepackage{natbib}
\usepackage{multirow}
\usepackage[usenames, dvipsnames,x11names]{xcolor}
\usepackage{hyperref}
\usepackage{bm,bbm}

\hypersetup{colorlinks=true, urlcolor=Turquoise4, linkcolor=Aquamarine, citecolor=Orchid4}
  \numberwithin{equation}{section}

\newcommand{\mathsym}[1]{{}}

\input{tcilatex}
\begin{document}

\title{A 3-3-1 model with low scale seesaw mechanisms}
\author{A. E. C\'{a}rcamo Hern\'{a}ndez}
\email{antonio.carcamo@usm.cl}
\author{Yocelyne Hidalgo Velásquez}
\email{yocehidalgov@gmail.com}
\author{Nicol\'{a}s A. P\'{e}rez-Julve}
\email{nicolasperezjulve@gmail.com}
\affiliation{Universidad T\'{e}cnica Federico Santa Mar\'{\i}a and Centro Cient\'{\i}%
fico-Tecnol\'{o}gico de Valpara\'{\i}so, \\
Casilla 110-V, Valpara\'{\i}so, Chile,}

\date{\today }

\begin{abstract}
We construct a viable 3-3-1 model with two $SU(3)_L$ scalar triplets, extended fermion and scalar spectrum, based on the $T^{\prime}$ family symmetry and other auxiliary cyclic symmetries, whose spontaneous breaking yields the observed pattern of SM fermion mass spectrum and fermionic mixing parameters. In our model the SM quarks lighter than the top quark, get their masses from a low scale Universal seesaw mechanism, the SM charged lepton masses are produced by a Froggatt-Nielsen mechanism and the small light active neutrino masses are generated from an inverse seesaw mechanism. The model is consistent with the low energy SM fermion flavor data and successfully accommodates the current Higgs diphoton decay rate and predicts charged lepton flavor violating decays within the reach of the forthcoming experiments.
\end{abstract}

\maketitle

\affiliation{Universidad T\'{e}cnica Federico Santa Mar\'{\i}a and Centro Cient\'{\i}fico-Tecnol\'{o}gico de Valpara\'{\i}so, \\
Casilla 110-V, Valpara\'{\i}so, Chile,}


\section{Introduction}

The existence of three fermion families and the observed pattern of Standard
Model (SM) fermion masses and mixing angles are not explained within the
context of the SM. Whereas in the quark sector, the mixing angles are small,
in the lepton sector two of the mixing angles are large and one is small, of
the order of the Cabbibo angle. The pattern of SM fermion masses is
extended over a range of 5 orders of magnitude in the quark sector and a
dramatically broader range of about 13 orders of magnitude, when the light
active neutrino sector is included. That flavour puzzle of the SM motivates
the study of theories with an extended particle spectrum and enlarged
symmetries, whose spontaneous breaking produces the observed SM fermion mass
and mixing hierarchy.

In addition, the SM predicts very tiny branching ratios for the charged
lepton flavor violating processes (cLFV) $\mu\to e\gamma$, $\tau\to\mu\gamma$
and $\tau\to e\gamma$, several orders of magnitude below their corresponding
projective experimental sensitivity. On the other low scale seesaw models 
\cite%
{Abada:2012cq,Deppisch:2004fa,Abada:2014kba,Abada:2014vea,Abada:2016awd,Abada:2018qok}
predict branching ratios for the cLFV processes within the reach of the
projective experimental sensitivity. Thus, a future observation of charged
lepton flavor violating processes will provide an undubitable evidence of
Physics Beyond the Standard Model and will shed light on the dynamics that
produces tiny light active neutrino masses and the measured leptonic mixing
angles.  

Furthermore, the origin of the family structure of fermions, which is not
addressed by the SM, can be explained in theories having an extended $%
SU(3)_C\times SU(3)_L\times U(1)_X$ gauge symmetry, called 3-3-1 models \cite{Georgi:1978bv,Valle:1983dk,Pisano:1991ee,Foot:1992rh,Frampton:1992wt,Hoang:1996gi, Hoang:1995vq,Foot:1994ym,CarcamoHernandez:2005ka,Dong:2010zu,Dong:2010gk,Dong:2011vb,Benavides:2010zw,Dong:2012bf,Huong:2012pg,Giang:2012vs,Binh:2013axa,Hernandez:2013mcf,Hernandez:2013hea,Hernandez:2014vta,Hernandez:2014lpa,Kelso:2014qka,Vien:2014gza,Phong:2014ofa,Phong:2014yca,Boucenna:2014ela,DeConto:2015eia,Boucenna:2015zwa,Boucenna:2015pav,Benavides:2015afa,Hernandez:2015tna,Hue:2015fbb,Hernandez:2015ywg,Fonseca:2016tbn,Fonseca:2016xsy,Deppisch:2016jzl,Reig:2016ewy,CarcamoHernandez:2017cwi,CarcamoHernandez:2017kra,Hati:2017aez,Barreto:2017xix,CarcamoHernandez:2018iel,Vien:2018otl,Dias:2018ddy,Ferreira:2019qpf,Huong:2019vej,CarcamoHernandez:2019iwh}. In these models, the cancellation of chiral anomalies takes place when the number of $SU(3)_L$ fermionic triplets is equal to the number of $SU(3)_L$
fermionic antitriplets, which happens when the number of fermion generations
is a multiple of three. Furthermore, when combined with the QCD asymptotic
freedom, the 3-3-1 models predict that the number of fermion generations is
exactly three. In addition, the nonuniversal $U(1)_X$ charge assignments for
the left handed quarks fields in the 3-3-1 models, are crucial for
explaining the large mass splitting between the heaviest quark and the two
lighter ones. Other phenomenological advantages of the 3-3-1 models are: 1)
they address the electric charge quantization \cite%
{deSousaPires:1998jc,VanDong:2005ux}, 2) they contain several sources of CP
violation \cite{Montero:1998yw,Montero:2005yb}, 3) they have a natural
Peccei-Quinn symmetry, thus allowing to address the strong-CP problem \cite%
{Pal:1994ba,Dias:2002gg,Dias:2003zt,Dias:2003iq}, 4) the 3-3-1 models with
heavy sterile neutrinos in the fermionic spectrum have cold dark matter
candidates as weakly interacting massive particles (WIMPs) \cite%
{Mizukoshi:2010ky,Dias:2010vt,Alvares:2012qv,Cogollo:2014jia}, 5) they
predict the bound $\sin\theta^2_W<\frac1{4}$, for the weak mixing parameter,
6) the 3-3-1 models with three right handed Majorana neutrinos and non SM
fermions without non SM electric charges, allow the implementation of a low
scale seesaw mechanism, which could be inverse or linear, thus allowing to
explain the smallness of the light active neutrinos masses and to predict
charged lepton flavor violating process within the reach of the forthcoming
experiments.

In this work, motivated by the aforementioned considerations, we propose a
3-3-1 model with two $SU(3)_L$ scalar triplets, extended fermion and scalar
spectrum, consistent with SM fermion masses and mixings. Our model
incorporates a Universal low scale seesaw mechanism to generate the masses for the SM
quarks lighter than the top quark, a Froggatt-Nielsen mechanism that
produces the SM charged lepton masses and an inverse seesaw mechanism that
gives rise to small light active neutrino masses. In our model we use the $%
T^{\prime}$ symmetry, which in combination with other auxiliary symmetries,
allows a viable description of the current SM fermion mass spectrum and
mixing parameters. We use the double tetrahedral group $T^{\prime }$ since it is the smallest discrete subgroup of $SU(2)$ as well as the smallest group of any kind with 1-, 2- and 3-dimensional representations and the multiplication rule $\mathbf{2}\otimes\mathbf{2}=\mathbf{3}\oplus\mathbf{1}$, thus allowing to reproduce the successful $U(2)$ textures \cite{Aranda:2000tm}. Note that the discrete group $T^{\prime }$ \cite%
{Aranda:2000tm,Feruglio:2007uu,Sen:2007vx,Aranda:2007dp,Chen:2007afa,Eby:2008uc,Frampton:2008bz,Frampton:2008ep,Eby:2009ii,Frampton:2009fw,Merlo:2011hw,Eby:2011ph,Eby:2011qa,Chen:2011tj,Meroni:2012ty,Frampton:2013lva,Chen:2013wba,Girardi:2013sza,Carone:2016xsi,Vien:2018otl,Carone:2019lfc}%
, together with the groups $A_{4}$ \cite%
{Ma:2001dn,He:2006dk,Feruglio:2008ht,Feruglio:2009hu,Chen:2009um,Varzielas:2010mp,Altarelli:2012bn,Ahn:2012tv,Memenga:2013vc,Felipe:2013vwa,Varzielas:2012ai, Ishimori:2012fg,King:2013hj,Hernandez:2013dta,Babu:2002dz,Altarelli:2005yx,Gupta:2011ct,Morisi:2013eca, Altarelli:2005yp,Kadosh:2010rm,Kadosh:2013nra,delAguila:2010vg,Campos:2014lla,Vien:2014pta,Joshipura:2015dsa,Hernandez:2015tna,Karmakar:2016cvb,Chattopadhyay:2017zvs,CarcamoHernandez:2017kra,Ma:2017moj,CentellesChulia:2017koy,Bjorkeroth:2017tsz,Srivastava:2017sno,Borah:2017dmk,Belyaev:2018vkl,CarcamoHernandez:2018aon,Srivastava:2018ser,delaVega:2018cnx,Borah:2018nvu,Pramanick:2019qpg,CarcamoHernandez:2019pmy,CarcamoHernandez:2019kjy,Ding:2019zxk}%
, $S_{4}$ \cite{Altarelli:2009gn,Bazzocchi:2009da,Bazzocchi:2009pv,Toorop:2010yh,Patel:2010hr,Morisi:2011pm,Altarelli:2012bn,Mohapatra:2012tb,BhupalDev:2012nm,Varzielas:2012pa,Ding:2013hpa,Ishimori:2010fs,Ding:2013eca,Hagedorn:2011un,Campos:2014zaa,Dong:2010zu,VanVien:2015xha,deAnda:2017yeb,deAnda:2018oik,CarcamoHernandez:2019eme,Chen:2019oey,deMedeirosVarzielas:2019cyj,deMedeirosVarzielas:2019hur,CarcamoHernandez:2019kjy,King:2019vhv,Chen:2019oey} and $\Delta (27)$ \cite%
{Branco:1983tn,deMedeirosVarzielas:2006fc,Ma:2007wu,Varzielas:2012nn,Bhattacharyya:2012pi,Ferreira:2012ri,Ma:2013xqa,Nishi:2013jqa,Varzielas:2013sla,Aranda:2013gga,Harrison:2014jqa,Ma:2014eka,Abbas:2014ewa,Abbas:2015zna,Varzielas:2015aua,Bjorkeroth:2015uou,Chen:2015jta,Vien:2016tmh,Hernandez:2016eod,CarcamoHernandez:2017owh,deMedeirosVarzielas:2017sdv,Bernal:2017xat,CarcamoHernandez:2018iel,deMedeirosVarzielas:2018vab,CarcamoHernandez:2018hst,CarcamoHernandez:2018djj,Bjorkeroth:2019csz}%
, is the smallest group containing an irreducible triplet representation that can accommodate the three fermion
families of the Standard model (SM). These groups have attracted a lot of attention of the model building community since they successfully describe the observed SM fermion mass spectrum and mixing parameters.  

The content of this paper goes as follows. In section \ref{model} we outline the proposed model, describing its fermionic and scalar spectrum as well as their assignments under the different continuous and discrete groups. The gauge sector of the model is discussed in section \ref{gaugesector}, whereas the scalar potential for two $SU(3)_L$ triplets is discussed in section  \ref{scalarpotentialtriplets}. The implications of our model in SM quark masses and mixings are discussed in section \ref{quarkmassesandmixing}. In Section \ref{leptonsector}, we present our results in terms of lepton masses and mixing, which is followed by a numerical analysis. The implications of our model in the Higgs diphoton decay rate are discussed in section \ref{htogammagamma}. In section \ref{LFV}, lepton flavor violating decays of the charged leptons are discussed, where sterile neutral lepton masses are constrained. Conclusions are given in section \ref{conclusions}. Some technical details are shown in the appendices: Appendix \ref{Tprime} provides a description of the $T^{\prime }$ discrete group. Appendix \ref{scalarpotentialTprime} includes a discussion of the scalar potential for a $T^{\prime }$ scalar triplet and its minimization condition.

\section{The model} 
\subsection{Particle content}
 \label{model} 
We consider a model based on the extended gauge symmetry $%
SU(3)_{C}\times SU(3)_{L}\times U(1)_{X}$ (3-3-1 model) which is
supplemented by the $U(1)_{L_{g}}$ global lepton number symmetry and the $T^{\prime }\times Z_{6}\times Z_{8}\times Z_{12}$ discrete group. Our model is an extension of the 3-3-1 model with two $SU(3)_{L}$ scalar triplets, where the scalar sector is augmented by the inclusion of several gauge singlet scalars and the fermion spectrum is enlarged by adding several vector like fermions and right handed Majorana neutrinos. The $SU(3)_{L}$ singlet vector like fermions are introduced in our model in order to implement a Universal Seesaw mechanism \cite{Davidson:1987mh,Berezhiani:1991ds,Sogami:1991yq} for the generation of the masses of SM quarks lighter than the top quark. We additionally introduce three gauge singlet right handed Majorana neutrinos which are crucial to incorporate the inverse seesaw mechanism in our model. In our model the non SM fermions do not have non SM electric charges, thus implying that the third component of the $SU(3)_L$ leptonic triplet is electrically neutral, thus allowing the implementation of an inverse seesaw mechanism \cite{Mohapatra:1986bd,GonzalezGarcia:1988rw,Akhmedov:1995vm,Akhmedov:1995ip,Malinsky:2005bi,Malinsky:2009df,Abada:2014vea} to generate the small light active neutrino masses. The SM charged lepton
masses are produced from a Froggatt-Nielsen mechanism \cite{Froggatt:1978nt}, which is triggered by non renormalizable Yukawa interactions involving the $SU(3)_{L}$ scalar triplets $\eta $ and $\chi $ as well as several gauge singlet scalars charged under the different discrete group factors of the model. In our model the hierarchy of SM charged fermion masses and fermionic mixing parameters is produced by the spontaneous breaking of the $T^{\prime }\times Z_{6}\times Z_{8}\times Z_{12} $ discrete group. The $SU(3)_{C}\times SU(3)_{L}\times U(1)_{X}\times T^{\prime }\times Z_{6}\times Z_{8}\times Z_{12}$ assignments of the scalar and fermionic fields of our model are shown in Tables \ref{tab:scalars} and \ref{tab:fermions}, respectively. Notice that in these tables the dimensions of the $SU(3)_C$, $SU(3)_L$ and $T^{\prime }$ representations are specified by the numbers
in boldface and the different $Z_{N}$ charges are written in additive
notation. Let us note that a field $\psi $ transforms under the $Z_{N}$ symmetry with a corresponding $q_{n}$ charge as: $\psi \rightarrow e^{\frac{2\pi iq_{n}}{N}}\psi $, $n=0,1,2,3\cdots N-1$. An explanation of the role of the different discrete group factors of the model is provided in the following. The double tetrahedral group $T^{\prime}$ selects the allowed entries of the mass matrices for SM charged fermions and neutrinos, thus allowing a reduction of the model parameters. In addition, as it will be shown below in Sections \ref{quarkmassesandmixing} and \ref{leptonsector}, the spontaneous breaking of the $T^{\prime}$ discrete group will be crucial to generate the observed CP violation in both quark and lepton sectors, without the need of invoking complex Yukawa couplings. Let us note that $T^{\prime }$ is the smallest discrete subgroup of $SU(2)$ as well as the smallest group of any kind with 1-, 2- and 3-dimensional representations and the multiplication rule $\mathbf{2}\otimes\mathbf{2}=\mathbf{3}\oplus\mathbf{1}$, thus allowing to reproduce the successful $U(2)$ textures as pointed out in Ref. \cite{Aranda:2000tm}. The $Z_6$ discrete group separates the $T^{\prime}$ scalar triplets ($\rho$, $\phi$ and $\zeta$) participating in the charged lepton Yukawa interactions from the one ($\xi$) appearing in the neutrino Yukawa terms, thus allowing to treat the charged lepton and neutrino sectors independently. The $Z_8$ discrete group contributes in generating small lepton number violating Majorana mass terms that yields a small $\mu$ parameter of the inverse seesaw mechanism that produces the tiny light active neutrino masses. Furthermore, $Z_8$ discrete group helps in shaping the texture for the SM charged leptons, that allows a reduction of the model parameters. The $Z_{12}$ discrete group is crucial for: 1) explaining the SM charged lepton mass hierarchy, 2) shaping the hierarchical structure of the quark mass matrices necessary to get a realistic pattern of quark masses and mixing and 3) generating small lepton number violating Majorana mass terms thus allowing to provide a natural explanation for the tiny values of the light active neutrino masses. %

The full symmetry $\mathcal{G}$ of our model features the following
two-step spontaneous breaking: 
\begin{eqnarray}
&&\mathcal{G}=SU(3)_{C}\times SU\left( 3\right) _{L}\times U\left( 1\right)
_{X}\times U(1)_{L_{g}}\times T^{\prime }\times Z_{6}\times Z_{8}\times
Z_{12}{\xrightarrow{\Lambda_{int}}}  \notag \\
&&\hspace{7mm}SU(3)_{C}\times SU\left( 2\right) _{L}\times U\left( 1\right)
_{Y}\times Z_{2}^{(L_{g})}{\xrightarrow{v_{\eta }}}  \notag \\
&&\hspace{7mm}SU(3)_{C}\times U\left( 1\right) _{Q}\times Z_{2}^{(L_{g})},
\label{SB}
\end{eqnarray}
where the symmetry breaking scales fulfill the hierarchy $\Lambda _{int}\sim v_{\chi }\gg v_{\eta }$. It is worth mentioning that the first step of symmetry breaking in Eq. (\ref{SB}) is triggered by the $SU(3)_L$ scalar triplet $\chi$, whose third component acquires a $10$ TeV scale vacuum expectation value (VEV) that breaks the $SU(3)_L\times U(1)_X$ gauge symmetry as well as by the $SU(3)_L$ scalar singlets whose VEVs break the $T^{\prime }\times Z_{6}\times Z_{8}\times Z_{12}$ discrete group. The non SM particles get masses at the $v_{\chi}$ scale after the spontaneous breaking of the $SU(3)_L\times U(1)_X$ gauge symmetry. We consider $v_{\chi }\sim \mathcal{O}(10)$ TeV, because the experimental data on $K$, $D$ and $B$ meson mixings set a lower bound of about $4$ TeV \cite{Huyen:2012uk} for the $Z^{\prime }$ gauge boson mass in 3-3-1 models, which translates in a lower limit of about $10$ TeV for the $SU\left( 3\right) _{L}\times U\left( 1\right) _{X}$ gauge symmetry breaking scale $v_{\chi }$. In addition, $v_{\chi }\sim \mathcal{O}(10)$ TeV is also consistent with the collider constraints as well as with the constraints that the decays $B_{s,d}\rightarrow \mu^{+}\mu ^{-}$ and $B_{d}\rightarrow K^{\ast }(K)\mu ^{+}\mu ^{-}$ impose on the $Z^{\prime }$ masses. It is worth mentioning that the LHC searches constrain the $Z^{\prime }$ gauge boson mass in 3-3-1 models to be larger than about $2.5$ TeV \cite{Salazar:2015gxa}, which corresponds to a lower limit of $6.3$ TeV for the  $SU(3)_{C}\times SU\left( 3\right) _{L}\times U\left( 1\right) _{X}$ symmetry breaking scale $v_{\chi }$. On the other hand, the decays $B_{s,d}\rightarrow \mu^{+}\mu ^{-}$ and $B_{d}\rightarrow K^{\ast }(K)\mu ^{+}\mu ^{-}$ set lower limits on the $Z^{\prime }$ gauge boson mass ranging from $1$ TeV up to $3$ TeV \cite{CarcamoHernandez:2005ka,Martinez:2008jj,Buras:2013dea,Buras:2014yna,Buras:2012dp}. Consequently, the scale $v_{\chi }\sim \mathcal{O}(10)$ TeV is consistent with the aforementioned constraints. Furthermore, we assume that the discrete symmetries of the model are broken at the same scale of breaking of the $SU(3)_L\times U(1)_X$ gauge symmetry. Moreover, let us note that the second step of symmetry breaking in Eq. (\ref{SB}) is triggered by the $SU(3)_L$ scalar triplet $\eta$, whose first component get a VEV that satisfies $v_{\eta}=v=246$ GeV and provides masses for the SM particles. Note that the $U(1)_{L_{g}}$ global lepton number symmetry is assumed to be spontaneously broken down to a residual discrete $Z_{2}^{(L_{g})}$ by the vacuum expectation value (VEV) of the $U(1)_{L_{g}}$ charged gauge-singlet scalar $\varphi $, having a nontrivial $U(1)_{L_{g}}$ charge, as indicated by Table \ref{tab:scalars}. The residual discrete $Z_{2}^{(L_{g})}$ lepton number symmetry, under which the leptons are charged and the other particles are neutral, forbids interactions involving an odd number of leptons, thus preventing proton decay. The corresponding massless Goldstone boson, namely, the Majoron, is phenomenologically harmless since it is a $SU(3)_{L}$ scalar singlet.  

Given that we are considering a 3-3-1 model where the non SM fermions do not have exotic electric charges, the electric charge in our model is defined in terms of the $SU(3)$ generators and the identity as follows:
\begin{equation}
Q=T_{3}+\beta T_{8}+XI=T_{3}-\frac{1}{\sqrt{3}}T_{8}+XI,
\end{equation}
with $I=diag(1,1,1)$, $T_{3}=\frac{1}{2}diag(1,-1,0)$ and $T_{8}=(\frac{1}{2 \sqrt{3}})diag(1,1,-2)$ for a $SU(3)_L$ triplet. Furthermore, the lepton number has a gauge component as well as a complementary global one, as indicated by the following relation:
\begin{equation}
L=\frac{4}{\sqrt{3}}T_{8}+L_{g},
\end{equation}
being $L_{g}$ a conserved charge associated with the $U(1)_{L_{g}}$ global lepton number symmetry.

The $SU(3)_L$ triplet scalar fields $\chi$ and $\eta$ can be expanded around the minimum as follows:
 \begin{eqnarray}
\chi  &=&%
\begin{pmatrix}
\chi _{1}^{0} \\ 
\chi _{2}^{-} \\ 
\frac{1}{\sqrt{2}}(v_{\chi }+\xi _{\chi }\pm i\zeta _{\chi })%
\end{pmatrix},\hspace{2cm}
\eta =%
\begin{pmatrix}
\frac{1}{\sqrt{2}}(v_{\eta }+\xi _{\eta }\pm i\zeta _{\eta }) \\ 
\eta _{2}^{-} \\ 
\eta _{3}^{0}%
\end{pmatrix}.%
\end{eqnarray}

The $SU(3)_L$ fermionic triplets and antitriplets can be represented as: 
\begin{eqnarray}
  Q_{nL}=
\begin{pmatrix}
D_{n} \\ 
-U_{n} \\ 
J_{n} \\ 
\end{pmatrix}_{L},
\hspace{1cm}
Q_{3L}=
\begin{pmatrix}
U_{3} \\ 
D_{3} \\ 
T \\ 
\end{pmatrix}_{L}, 
\hspace{1cm}
L_{iL}=%
\begin{pmatrix}
\nu _{i} \\ 
e_{i} \\ 
\nu _{i}^{c} \\ 
\end{pmatrix}%
_{L},\hspace{1cm}n=1,2,\hspace{1cm}i=1,2,3.
\end{eqnarray}
\begin{table}[th]
\begin{tabular}{|c|c|c|c|c|c|c|c|c|c|c|c|c|c|c|}
\hline
& $\chi $ & $\eta $ & $\varphi $ & $\sigma $ & $\xi $ & $\rho $ & $\phi $ & $%
\zeta $ & $S_{1}$ & $S_{2}$ & $S_{3}$ & $S_{4}$ & $S_{5}$ & $S_{6}$ \\ \hline
$SU(3)_{C}$ & $\mathbf{1}$ & $\mathbf{1}$ & $\mathbf{1}$ & $\mathbf{1}$ & $%
\mathbf{1}$ & $\mathbf{1}$ & $\mathbf{1}$ & $\mathbf{1}$ & $\mathbf{1}$ & $%
\mathbf{1}$ & $\mathbf{1}$ & $\mathbf{1}$ & $\mathbf{1}$ & $\mathbf{1}$ \\ 
\hline
$SU(3)_{L}$ & $\mathbf{3}$ & $\mathbf{3}$ & $\mathbf{1}$ & $\mathbf{1}$ & $%
\mathbf{1}$ & $\mathbf{1}$ & $\mathbf{1}$ & $\mathbf{1}$ & $\mathbf{1}$ & $%
\mathbf{1}$ & $\mathbf{1}$ & $\mathbf{1}$ & $\mathbf{1}$ & $\mathbf{1}$ \\ 
\hline
$U(1)_{X}$ & $-\frac{1}{3}$ & $-\frac{1}{3}$ & $0$ & $0$ & $0$ & $0$ & $0$ & 
$0$ & $0$ & $0$ & $0$ & $0$ & $0$ & $0$ \\ \hline
$U(1)_{L_{g}}$ & $\frac{4}{3}$ & $-\frac{2}{3}$ & $2$ & $0$ & $0$ & $0$ & $0$
& $0$ & $0$ & $0$ & $0$ & $0$ & $0$ & $0$ \\ \hline
$T^{\prime }$ & $\mathbf{1}$ & $\mathbf{1}$ & $\mathbf{1}^{\prime }$ & $\mathbf{1}$ & $%
\mathbf{3}$ & $\mathbf{3}$ & $\mathbf{3}$ & $\mathbf{3}$ & $\mathbf{2}$ & $%
\mathbf{2}^{\prime }$ & $\mathbf{2}^{\prime }$ & $\mathbf{1}$ & $\mathbf{1}%
^{\prime }$ & $\mathbf{1}^{\prime \prime }$ \\ \hline
$Z_{6}$ & $0$ & $0$ & $0$ & $0$ & $0$ & $2$ & $2$ & $2$ & $0$ & $0$ & $0$ & $%
3$ & $0$ & $0$ \\ \hline
$Z_{8}$ & $0$ & $0$ & $-4$ & $0$ & $0$ & $-1$ & $-2$ & $0$ & $0$ & $0$ & $0$
& $0$ & $1$ & $1$ \\ \hline
$Z_{12}$ & $0$ & $0$ & $-4$ & $-1$ & $0$ & $0$ & $0$ & $0$ & $0$ & $0$ & $1$
& $0$ & $-1$ & $0$ \\ \hline
\end{tabular}%
\caption{Scalar assignments under $SU(3)_{C}\times SU(3)_{L}\times U(1)_{X}\times T^{\prime }\times Z_{6}\times Z_{8}\times
Z_{12}$.}
\label{tab:scalars}
\end{table}

\begin{table}[th]
\begin{tabular}{|c|c|c|c|c|c|c|c|c|c|c|c|c|c|c|c|c|c|c|c|c|c|c|c|}
\hline
& $Q_{L}$ & $Q_{3L}$ & $U_{1R}$ & $U_{2R}$ & $U_{3R}$ & $D_{1R}$ & $D_{2R}$
& $D_{3R}$ & $T_{R}$ & $J_{R}$ & $\widetilde{T}_{L}$ & $\widetilde{T}_{R}$ & 
$B_{1L}$ & $B_{1R}$ & $B_{2L}$ & $B_{2R}$ & $B_{3L}$ & $B_{3R}$ & $L_{L}$ & $%
e_{1R}$ & $e_{2R}$ & $e_{3R}$ & $N_{R}$ \\ \hline
$SU(3)_{C}$ & $\mathbf{3}$ & $\mathbf{3}$ & $\mathbf{3}$ & $\mathbf{3}$ & $%
\mathbf{3}$ & $\mathbf{3}$ & $\mathbf{3}$ & $\mathbf{3}$ & $\mathbf{3}$ & $%
\mathbf{3}$ & $\mathbf{3}$ & $\mathbf{3}$ & $\mathbf{3}$ & $\mathbf{3}$ & $%
\mathbf{3}$ & $\mathbf{3}$ & $\mathbf{3}$ & $\mathbf{3}$ & $\mathbf{1}$ & $%
\mathbf{1}$ & $\mathbf{1}$ & $\mathbf{1}$ & $\mathbf{1}$ \\ \hline
$SU(3)_{L}$ & $\mathbf{3^{\ast }}$ & $\mathbf{3}$ & $\mathbf{1}$ & $\mathbf{1%
}$ & $\mathbf{1}$ & $\mathbf{1}$ & $\mathbf{1}$ & $\mathbf{1}$ & $\mathbf{1}$
& $\mathbf{1}$ & $\mathbf{1}$ & $\mathbf{1}$ & $\mathbf{1}$ & $\mathbf{1}$ & 
$\mathbf{1}$ & $\mathbf{1}$ & $\mathbf{1}$ & $\mathbf{1}$ & $\mathbf{1}$ & $%
\mathbf{1}$ & $\mathbf{1}$ & $\mathbf{1}$ & $\mathbf{1}$ \\ \hline
$U(1)_{X}$ & $0$ & $\frac{1}{3}$ & $\frac{2}{3}$ & $\frac{2}{3}$ & $\frac{2}{%
3}$ & $-\frac{1}{3}$ & $-\frac{1}{3}$ & $-\frac{1}{3}$ & $\frac{2}{3}$ & $-%
\frac{1}{3}$ & $\frac{2}{3}$ & $\frac{2}{3}$ & $-\frac{1}{3}$ & $-\frac{1}{3}
$ & $-\frac{1}{3}$ & $-\frac{1}{3}$ & $-\frac{1}{3}$ & $-\frac{1}{3}$ & $-%
\frac{1}{3}$ & $-1$ & $-1$ & $-1$ & $0$ \\ \hline
$U(1)_{L_{g}}$ & $\frac{2}{3}$ & $-\frac{2}{3}$ & $0$ & $0$ & $0$ & $0$ & $0$
& $0$ & $-2$ & $2$ & $0$ & $0$ & $0$ & $0$ & $0$ & $0$ & $0$ & $0$ & $\frac{1%
}{3}$ & $1$ & $1$ & $1$ & $-1$ \\ \hline
$T^{\prime }$ & $\mathbf{2}$ & $\mathbf{1}^{\prime \prime }$ & $\mathbf{1}$
& $\mathbf{1}^{\prime }$ & $\mathbf{1}^{\prime \prime }$ & $\mathbf{1}$ & $%
\mathbf{1}^{\prime }$ & $\mathbf{1}^{\prime \prime }$ & $\mathbf{1}^{\prime
\prime }$ & $\mathbf{2}$ & $\mathbf{2}$ & $\mathbf{2}$ & $\mathbf{1}$ & $%
\mathbf{1}$ & $\mathbf{1}^{\prime }$ & $\mathbf{1}^{\prime }$ & $\mathbf{1}%
^{\prime \prime }$ & $\mathbf{1}^{\prime \prime }$ & $\mathbf{3}$ & $\mathbf{%
1}^{\prime }$ & $\mathbf{1}^{\prime }$ & $\mathbf{1}$ & $\mathbf{3}$
\\ \hline
$Z_{6}$ & $0$ & $0$ & $3$ & $3$ & $0$ & $-3$ & $-3$ & $-3$ & $0$ & $0$ & $3$
& $0$ & $0$ & $0$ & $0$ & $0$ & $0$ & $0$ & $0$ & $-2$ & $-2$ & $-2$ & $0$
\\ \hline
$Z_{8}$ & $0$ & $0$ & $0$ & $-1$ & $0$ & $0$ & $0$ & $0$ & $0$ & $0$ & $0$ & 
$0$ & $0$ & $0$ & $0$ & $0$ & $0$ & $0$ & $0$ & $0$ & $2$ & $-1$ & $0$ \\ 
\hline
$Z_{12}$ & $0$ & $0$ & $5$ & $1$ & $0$ & $2$ & $2$ & $2$ & $0$ & $0$ & $0$ & 
$0$ & $2$ & $2$ & $0$ & $0$ & $0$ & $0$ & $-1$ & $5$ & $2$ & $-1$ & $-1$ \\ 
\hline
\end{tabular}%
\caption{Fermion assignments under $SU(3)_{C}\times SU(3)_{L}\times U(1)_{X}\times T^{\prime }\times Z_{6}\times
Z_{8}\times Z_{12}$.}
\label{tab:fermions}
\end{table}
With the particle content shown in Tables \ref{tab:scalars} and \ref{tab:fermions}, the following relevant Yukawa terms for the quark and lepton sector invariant under the group $\mathcal{G}$ arise:
\begin{eqnarray}
-\mathcal{L}_{Y}^{\left( q\right) } &=&y^{\left( T\right) }\overline{Q}%
_{3L}\chi T_{R}+y^{\left( J\right) }\left( \overline{Q}_{L}\chi ^{\ast
}J_{R}\right) _{\mathbf{1}}+y_{3}^{\left( U\right) }\overline{Q}_{3L}\eta
U_{3R}+y_{\widetilde{T}}\left( \overline{\widetilde{T}}_{L}S_{4}\widetilde{T}%
_{R}\right) _{\mathbf{1}}+\sum_{j=1}^{3}\left( m_{B}\right) _{j}\overline{B}%
_{jL}B_{jR}  \notag \\
&&+x_{11}^{\left( U\right) }\overline{\widetilde{T}}_{L}S_{1}U_{1R}\frac{%
\sigma ^{5}}{\Lambda ^{5}}+x_{22}^{\left( U\right) }\overline{\widetilde{T}}%
_{L}S_{2}U_{2R}\frac{S_{5}}{\Lambda }+x_{12}^{\left( U\right) }\overline{%
\widetilde{T}}_{L}S_{1}U_{2R}\frac{S_{6}\sigma }{\Lambda ^{2}}+y^{\left(
U\right) }\varepsilon _{abc}\left( \overline{Q}_{L}^{a}\eta ^{b}\chi ^{c}%
\widetilde{T}_{R}\right) _{\mathbf{1}}\frac{1}{\Lambda }  \notag \\
&&+x_{j}^{\left( D\right) }\overline{B}_{jL}S_{4}D_{jR}\frac{\sigma ^{2}}{%
\Lambda ^{2}}+y_{11}^{\left( D\right) }\overline{Q}_{L}\eta ^{\ast }B_{1R}%
\frac{S_{1}\sigma ^{2}}{\Lambda ^{3}}+y_{22}^{\left( D\right) }\overline{Q}%
_{L}\eta ^{\ast }B_{2R}\frac{S_{2}^{\ast }}{\Lambda }+y_{23}^{\left(
D\right) }\overline{Q}_{L}\eta ^{\ast }B_{3R}\frac{S_{2}}{\Lambda }  \notag
\\
&&+y_{13}^{\left( D\right) }\overline{Q}_{L}\eta ^{\ast }B_{3R}\frac{%
S_{3}\sigma }{\Lambda ^{2}}+y_{33}^{\left( D\right) }\varepsilon _{abc}%
\overline{Q}_{3L}^{a}\left( \eta ^{\ast }\right) ^{b}\left( \chi ^{\ast
}\right) ^{c}B_{3R}\frac{1}{\Lambda }+H.c,  \label{eqn:Lyq}
\end{eqnarray}%
\begin{eqnarray}
-\mathcal{L}_{Y}^{\left( l\right) } &=&y_{1}^{\left( L\right) }\varepsilon
_{abc}\left( \overline{L}_{L}^{a}\left( \eta ^{\ast }\right) ^{b}\left( \chi
^{\ast }\right) ^{c}\rho \right) _{\mathbf{\mathbf{1}}}e_{1R}\frac{\sigma
^{6}S_6}{\Lambda^{9}}+y_{2}^{\left( L\right) }\varepsilon _{abc}\left( 
\overline{L}_{L}^{a}\left( \eta ^{\ast }\right) ^{b}\left( \chi ^{\ast
}\right) ^{c}\phi \right) _{\mathbf{1}^{\prime \prime }}e_{2R}\frac{\sigma
^{3}}{\Lambda ^{5}}+\frac{y_{3}^{\left( L\right) }}{\Lambda ^{2}}\varepsilon
_{abc}\left( \overline{L}_{L}^{a}\left( \eta ^{\ast }\right) ^{b}\left( \chi
^{\ast }\right) ^{c}\zeta \right) _{\mathbf{1}^{\prime }}e_{3R}\frac{S_6}{\Lambda}\notag \\
&&+y_{\rho}\varepsilon _{abc}\varepsilon _{dec}\left( \overline{L}%
_{L}^{a}\left( L_{L}^{C}\right) ^{b}\right) _{\mathbf{3}_{2}}\eta ^{d}\chi
^{e}\frac{\xi \sigma ^{2}}{\Lambda ^{4}}+y_{\chi }^{\left( L\right) }\left( 
\overline{L}_{L}\chi N_{R}\right) _{\mathbf{\mathbf{1}}}\notag \\
&&+h_{1N}\left( N_{R}\overline{N_{R}^{C}}\right) _{\mathbf{\mathbf{1}}%
}\varphi \frac{\left( \sigma ^{\ast }\right)^{6}S^{4}_6}{\Lambda^{10}}+h_{2N}\left( N_{R}\overline{N_{R}^{C}}\right) _{\mathbf{3}_{1}}\varphi 
\frac{\xi\left( \sigma ^{\ast }\right)^{6}S^{4}_6}{\Lambda ^{11}}+H.c.,
\label{eqn:lyl}
\end{eqnarray}
where the dimensionless couplings in Eqs. (\ref{eqn:Lyq}) and (\ref{eqn:lyl}) are $\mathcal{O}(1)$ parameters.

As shown in detail in the Appendix \ref{scalarpotentialTprime}, the following VEV patterns for the $T^{\prime }$ scalar triplets are consistent with the scalar potential
minimization equations for a large region of parameter space: 
\begin{eqnarray}\
\left\langle \rho \right\rangle  &=&v_{\rho }\left( e^{-i\alpha }\left( \cos
\gamma -e^{i\left( 2\phi _{1}+\phi _{2}\right) }\sin \gamma \right)
,1,e^{i\alpha }\left( \cos \gamma -e^{-i\left( 2\phi _{1}+\phi _{2}\right)
}\sin \gamma \right) \right) ,  \notag \\
\left\langle \phi \right\rangle  &=&v_{\phi }\left( 1,e^{i\alpha }\left(
\cos \gamma +e^{i\left( \phi _{2}-\phi _{1}\right) }\sin \gamma \right)
,e^{-i\alpha }\left( \cos \gamma +e^{-i\left( \phi _{2}-\phi _{1}\right)
}\sin \gamma \right) \right) ,  \notag \\
\left\langle \zeta \right\rangle  &=&v_{\zeta }\left( e^{i\alpha }\left(
\cos \gamma +\sin \gamma \right) ,e^{-i\alpha }\left( \cos \gamma +\sin
\gamma \right) ,1\right),\hspace{1cm}\left\langle \xi \right\rangle =\frac{v_{\xi }}{\sqrt{3}}\left( 1,1,1\right).
\label{eqn:vevt}
\end{eqnarray}
In what regards the $T^{\prime }$ scalar doublets, we consider the following VEV configurations, which are natural solutions of the scalar potential minimization conditions:
\begin{eqnarray}
\left\langle S_{1}\right\rangle  &=&v_{S_{1}}\left( 0,1\right) ,\hspace{1cm}%
\left\langle S_{2}\right\rangle =v_{S_{2}}\left( -1,0\right) ,\hspace{1cm}%
\left\langle S_{3}\right\rangle =v_{S_{3}}\left( 0,1\right).
\label{eqn:vevt2}
\end{eqnarray}
Furthermore, since the observed pattern of the SM charged fermion masses and quark mixing angles is produced by the spontaneous breaking of the $T^{\prime }\times Z_{6}\times Z_{8}\times Z_{12}$ discrete group, we set the VEVs of the $SU(3)_{L}$ singlet scalar fields with respect to the Wolfenstein parameter $\lambda =0.225$ and the model cutoff $\Lambda $, as follows:
\begin{equation}
v_{S_{k}}\sim v_{\eta }\sim \lambda ^{3}\Lambda\ll v_{\chi }\sim v_{\zeta }\sim v_{\rho}\sim v_{\varphi }\sim v_{\sigma }\sim
v_{\xi }\sim v_{\phi }\sim v_{S_{l}}\sim \Lambda _{int}=\lambda \Lambda ,%
\hspace{0.5cm}k=1,2,3,\hspace{0.5cm}l=4,5,6.  \label{VEVsinglets}
\end{equation}
Considering $\Lambda_{int}\sim\mathcal{O}(10)$ TeV, from Eq. (\ref{VEVsinglets}) we find for the model cutoff the estimate $\Lambda\sim\mathcal{O}(40)$ TeV.
\subsection{The gauge sector}
\label{gaugesector}
The gauge bosons associated with the group $SU(3)_{L}$ for the case $\beta=-1/\sqrt{3}$ are written as follows:
\begin{align}
\mathbf{W}_{\mu }& =W_{\mu }^{\alpha }G_{\alpha }  \notag \\
& =\frac{1}{2}%
\begin{pmatrix}
W_{\mu }^{3}+\frac{1}{\sqrt{3}}W_{\mu }^{8} & \sqrt{2}W_{\mu }^{+} & \sqrt{2}%
K_{\mu }^{0} \\ 
\sqrt{2}W_{\mu }^{-} & -W_{\mu }^{3}+W_{\mu }^{8} & \sqrt{2}K_{\mu }^{-} \\ 
\sqrt{2}\overline{K}_{\mu }^{0} & \sqrt{2}K_{\mu }^{+} & -\frac{2}{\sqrt{3}}%
W_{\mu }^{8}%
\end{pmatrix},
\label{eqn:w}
\end{align}
where the electric charges of each gauge field correspond to the entries of the matrix: 
\begin{align}
Q_W =& 
\begin{pmatrix}
0 & 1 & 0 \\ 
-1 & 0 & -1 \\ 
0 & 1 & 0%
\end{pmatrix}.%
\end{align}
The gauge field associated with the $U(1)_X$ symmetry is electrically neutral, i.e., it has $Q_{B}=0$ and is represented as follows: 
\begin{align}
\mathbf{B}_{\mu} = \mathbf{I}_{3\times3}B_{\mu}.
\end{align}
The gauge sector associated with the $SU(3)_L\times U(1)_X$ group of the 3-3-1 models, is composed of five electrically neutral and four electrically charged gauge bosons. In the gauge boson spectrum there is one massless electrically neutral gauge boson which corresponds to the photon and eight massive gauge boson fields, namely, $Z$, $W^{\pm}$, $Z^{\prime }$, $W^{\prime \pm}$, $\overline{K}^{0}$, $K^{0}$. Five of the massive gauge bosons, $Z^{\prime }$, $W^{\prime \pm}$, $\overline{K}^{0}$, $K^{0}$ acquire their masses after the spontaneous breaking of the $SU(3)_L\times U(1)_X$ gauge symmetry down to the $SU(2)_L\times U(1)_Y$, whereas the $Z$ and $W^{\pm}$ gauge bosons become massive after electroweak symmetry breaking. 

The gauge boson mass terms as well as interactions between the scalar and gauge bosons arise from the following kinetic term: 
\begin{align}
\mathcal{L}_K=&\sum_{\Phi=\eta,\chi} (D^{\mu}\Phi)^{\dagger}(D_{\mu}\Phi)  \notag \\
=&\sum_{\Phi=\eta,\chi}\left[\overbrace{(\partial^{\mu}\Phi)^{\dagger}(D_{\mu}\Phi)+(D^{\mu}\Phi)^{%
\dagger}(\partial_{\mu}\Phi)}^{(1)}-(\partial^{\mu}\Phi)^{\dagger}(%
\partial_{\mu}\Phi) + \overbrace{\Phi^{\dagger}(gW^{\mu}+g^{\prime
}X_{\Phi}B^{\mu})^{\dagger}(gW^{\mu}+g^{\prime }X_{\Phi}B^{\mu})\Phi}^{(2)}\right],
\label{eqn:lagrangian}
\end{align}
where the covariant derivative in 3-3-1 models is defined as follows \cite{Diaz:2003dk}:
\begin{equation}
D_{\mu} = \partial_{\mu} + igW^{\alpha}_{\mu}G_{\alpha} + ig^{\prime
}X_{\Phi}B_{\mu}.\label{eqn:covder}
\end{equation}
Notice that the first two terms of Eq. (\ref{eqn:lagrangian}), which are denoted as $(1)$, include the couplings between the gauge bosons and the derivatives of the scalar fields, thus allowing to get information
about each would-be Goldstone boson interacting with its corresponding massive gauge boson. In addition, the last term of Eq. (\ref{eqn:lagrangian}), which is denoted as $(2)$, contains information about the masses of the gauge bosons and its couplings with the physical scalar fields.

The different entries of the gauge boson squared mass matrices are obtained from the following relation:
\begin{align}
M^{2}_{V_iV_j} = \frac{\partial^{2} \mathcal{L}_K}{\partial V_i
\partial V_j},
\end{align}
where for the charged gauge bosons $V_i = W^{\pm},W^{\prime \pm}$, whereas for the neutral ones $V_i = W^{3}, W^{8}, B, \overline{K}^{0}, K^{0}$.
Then, the squared mass matrices for the charged and neutral gauge bosons are respectively given by:
\begin{gather}
M^{2}_{charged} = \left( 
\begin{array}{cc}
\frac{1}{4} g^2 v_{\eta}^2 & 0 \\ 
0 & \frac{1}{4} g^2 v_{\chi}^2 
\end{array}
\right), \\
M^2_{neutral} = \left( 
\begin{array}{cccc}
\frac{1}{4} g^2 v_{\eta}^2 & \frac{g^2 v_{\eta}^2%
}{4 \sqrt{3}} & -\frac{1}{6} g v_{\eta}^2 g^{\prime } & 0 \\ 
\frac{g^2 v_{\eta}^2}{4 \sqrt{3}} & \frac{1}{12} g^2 v_{\eta}^2+\frac{1}{3} g^2 v_{\chi}^2 & \frac{g v_{\chi}^2 g^{\prime}}{3 \sqrt{3}}-\frac{g v_{\eta}^2
g^{\prime}}{6 \sqrt{3}} & 0 \\ 
-\frac{1}{6} g v_{\eta}^2 g^{\prime} & \frac{g v_{\chi}^2 g^{\prime }}{3 \sqrt{3}}-\frac{g v_{\eta}^2
g^{\prime }}{6 \sqrt{3}} & \frac{1}{9} v_{\eta}^2
\left(g^{\prime }\right)^2+\frac{1}{9} v_{\chi}^2
\left(g^{\prime }\right)^2 & 0 \\ 
0 & 0 & 0 & \frac{1}{8} g^2 v_{\eta}^2+\frac{1}{8} g^2 v_{\chi}^2 
\end{array}
\right).
\end{gather} 
The gauge bosons mass spectrum of the model is summarized in Table \eqref{tab:gaugebosons}\newline
\begin{minipage}{0.45\textwidth}
\begin{table}[H]
\centering
\renewcommand{\arraystretch}{1.4}
\begin{tabular}{c|c}
\textbf{Gauge Boson} & \textbf{Squared Mass}  \\
\hline\hline\
$W^{\pm}$ & $\frac{1}{4}g^2 v_{\eta}^2$ \\
\hline
$W'^{\pm}$ & $\frac{1}{4}g^2 v_{\chi}^2$ \\
\hline
$\gamma$ & $0$  \\ 
\hline
$Z$ & $\frac{1}{9} \left( \Xi_1 - \Xi_2 \right)$  \\ 
\hline
$Z^{\prime }$ & $\frac{1}{9} \left(  \Xi_1 + \Xi_2  \right)$ \\ 
\hline
$K^{0}, \overline{K}^{0}$ & $\frac{g^2}{8} \left(  v_{\chi}^2  + v_{\eta}^2  \right)$ \\ 
\hline 
\end{tabular}
\caption{Physical gauge bosons mass spectrum.}
\label{tab:gaugebosons}
\end{table}
\end{minipage}
\begin{minipage}{4\textwidth}
$\Xi_1 = 3 g^2 (v_{\eta}^2 + v_{\chi}^2)+(g^{\prime})^2 \left(v_{\eta}^2+v_{\chi}^2\right)$, \\ 
$\Xi_2=\sqrt{\left(3 g^2+\left(g^{\prime
}\right)^2\right)^2 \left(v_{\eta}^2+v_{\chi}^2\right)^{2} - 9 g^2 \left(3 g^2+4 \left(g^{\prime }\right)^2\right) v_{\eta}^2 v_{\chi}^2}$.
\end{minipage}
The squared masses for the $Z$ and $Z'$ gauge bosons can be approximatelly written as 
\cite{Hoang:1995vq}:
\begin{align}
    M^{2}_{Z} &= \frac{g^2}{4c_{W}}v^2_{\eta}, \label{eqn:mz}\\
    M^{2}_{Z'} &= \frac{g^2}{3-4s^{2}_{W}}v^2_{\chi}, \label{eqn:mzp}
\end{align}
where $c_{W} = \cos \theta_{W}$, $s_{W} = \sin \theta_{W}$ and $v_{\eta}=246$ GeV. Consequently, for $v_{\chi}\approx 10$ TeV, we find that the heavy gauge bosons have the masses $M_{W^{\prime}}\approx 3.3$ TeV and $M_{Z^{\prime }}\approx 4.5$ TeV.

\subsection{Scalar potential for two $SU(3)_{L}$ scalar triplets}
\label{scalarpotentialtriplets}
For the sake of simplicity we neglect the mixing terms between the $SU(3)_L$ scalar triplets and the gauge singlet scalars. Then, the 
scalar potential for two $SU(3)_{L}$ scalar triplets takes the form:
\begin{equation}
V = -\mu _{\chi}^{2}(\chi ^{\dagger }\chi )-\mu _{\eta }^{2}(\eta ^{\dagger
}\eta ) +\lambda _{1}(\chi^{\dagger }\chi )(\chi ^{\dagger }\chi )+\lambda
_{2}(\eta ^{\dagger }\eta )(\eta ^{\dagger}\eta ) +\lambda_{3}(\chi
^{\dagger }\chi )(\eta ^{\dagger }\eta )+\lambda _{4}(\chi ^{\dagger }\eta
)(\eta ^{\dagger }\chi ),
\label{eqn:scalpot}
\end{equation}
where $\chi$ and $\eta$ are the $SU(3)_L$ scalar triplets acquiring vacuum expectation values (VEVs) in their third and
 first components, respectively. The minimization conditions of the aforementioned scalar potential yields the following relations:
\begin{align}
    \frac{\partial V}{\partial v_{\chi}} = \frac{1}{2} \lambda _3 v _{\eta}^2 v _{\chi}+\lambda _1 v _{\chi}^3-\mu _1^2 v _{\chi} = 0, \\
   \frac{\partial V}{\partial v_{\eta}} = \frac{1}{2} \lambda _3 v _{\eta } v _{\chi}^2+\lambda _2 v _{\eta}^3-\mu _2^2 v _{\eta} = 0.
\end{align}
Thus, the VEV patterns for the $SU(3)_L$ scalar triplets $\chi$ and $\eta$ are compatible with a global minimum of the scalar potential of \eqref{eqn:scalpot}.
Solving these equations, the mass parameters can be obtained: 
\begin{align}
\mu_\chi^2 &= \frac{1}{2} \lambda_3 v_{\eta}^2 + \lambda_1 v_{\chi}^2, \\
\mu_\eta^2 &= \frac{1}{2} \lambda_3 v_{\chi}^2 + \lambda_2 v_{\eta}^2,
\end{align}
replacing these mass parameters in the Higgs potential, the neutral and
charged scalar mass spectrum resulting from the two $SU(3)_L$ scalar triplets can be obtained from the following relations:
\begin{align}
M^2_{\Phi_{i} \Phi_{j}} &= \frac{\partial^2 V}{\partial \Phi_{i}\Phi_{j}}%
\biggr|_{\Phi_i = 0}, & M^2_{\Phi^{*}_{i} \Phi_{j}} &=\frac{\partial^2 V}{%
\partial \Phi_{i}\Phi_{j}}\biggr|_{\Phi_i = 0},
\end{align}
for the neutral scalar masses $\Phi_{i}=\xi_{\chi},\xi_{\eta},\zeta_{\chi},%
\zeta_{\eta},\chi^{0},\eta^{0}$ and charged scalar masses $
\Phi_{i}=\chi^{\pm},\eta^{\pm}$ respectively. The scalar mass matrices are shown below: 
\begin{align*}
M^2_{\zeta\zeta} = 0_{2\times2}, & & M^2_{\chi^{\pm} \eta^{\pm}} =
0_{2\times2},
\end{align*}
\begin{align}
M^2_{\chi^0 \eta^0} = \left( 
\begin{array}{cc}
\lambda _4 v _{\text{$\eta $}}^2 & \lambda _4 v _{\text{$\eta $}} v _{%
\text{$\chi $}} \\ 
\lambda _4 v _{\text{$\eta $}} v _{\text{$\chi $}} & \lambda _4 v _{%
\text{$\chi $}}^2 
\end{array}
\right), & & M^2_{\xi\xi} = \left( 
\begin{array}{cc}
2 \lambda _1 v _{\text{$\chi $}}^2 & \lambda _3 v _{\text{$\eta $}} v
_{\text{$\chi $}} \\ 
\lambda _3 v _{\text{$\eta $}} v _{\text{$\chi $}} & 2 \lambda _2 v _{%
\text{$\eta $}}^2 
\end{array}
\right).
\end{align}
Finally, the physical scalar mass spectrum resulting from the $SU(3)_L$ scalar triplets $\eta$ and $\chi$ is summarized in Table \ref{tab:escalares}. 
\begin{minipage}{0.55\textwidth}
\begin{table}[H]
\centering
\renewcommand{\arraystretch}{1.3}
\begin{tabular}{c|c}
\textbf{Scalars} & \textbf{Masses}  \\
\hline\hline\
$G_1^0 = \zeta_\chi$ & $M^2_{G_1^0}=0$ \\
\hline
$G_2^0 = \zeta_\eta$ & $M^2_{G_2^0}=0$  \\
\hline
$h^0_1 = C_\alpha \xi_\chi - S_\alpha \xi_\eta$  &  $M^2_{h^0_1}=\Delta_1-\Delta_2$  \\ 
\hline
$H^0_1 = S_\alpha \xi_\chi + C_\alpha \xi_\eta$  & $M^2_{H^0_1}=\Delta_1+\Delta_2$ \\ 
\hline
$G_3^0 = -C_\beta \chi^0 + S_\beta \eta^0$  &  $M^2_{G_3^0}= 0$  \\ 
\hline
$H^0_2 = S_\beta \chi^0 + C_\beta \eta^0$  & $M^2_{H^0_2}=\lambda_4(v^2_\eta+v^2_\chi)$ \\ 
\hline
$G^{\pm}_1 = \chi^{\pm}$  &  $M^2_{G^{\pm}_1}=0$ \\ 
\hline
$G^{\pm}_2 = \eta^{\pm}$  & $M^2_{G^{\pm}_2}= 0$ \\ 
\end{tabular}
\caption{Physical scalar mass spectrum.}
\label{tab:escalares}
\end{table}
\end{minipage}
\begin{minipage}{4\textwidth}
$\Delta_1=\lambda_2 v_{\eta }^2+\lambda_1 v_{\chi }^2$, \\ $\Delta_2=\sqrt{\lambda_3^2 v_{\eta }^2 v_{\chi }^2-2 \lambda_1 \lambda_2 v_{\eta }^2 v _{\chi }^2+\lambda_2^2 v_{\eta }^4+\lambda_1^2 v_{\chi }^4}$,\\ $\tan(2\alpha)=\dfrac{\lambda_3 v_\eta v_\chi}{\lambda_1 v_\chi^2 -\lambda_2 v_\eta^2}$, \\  $\tan(\beta)=\dfrac{v_\chi}{v_\eta}$.
\end{minipage}
The physical scalar spectrum resulting from the two $SU(3)_L$ scalar triplets is composed of the
following fields: 2 CP-even Higgs bosons $(h_1^0, H_1^0)$ and one neutral
Higgs boson $(H_2^0)$. The scalar $h_1^0$ is identified with the SM-like $%
125$ GeV Higgs boson found at the LHC. It's noteworthy that the neutral
Goldstone bosons $G_1^0$, $G_2^0$, $G_3^0$ and $\overline{G}_3^0$ are
associated to the longitudinal components of the $Z$, $Z^{\prime }$, $K^0$
and $\overline{K}^0$ gauge bosons. Furthermore, the charged Goldstone bosons $G^{\pm}_1$
and $G^{\pm}_2$ are associated to the longitudinal components of the $W^\pm$
and $W'^\pm$ gauge bosons respectively.

Finally to close this section, we briefly comment about the LHC signals of a $Z^{\prime}$ gauge boson. The heavy $Z^{\prime}$ gauge boson is mainly produced via Drell-Yan mechanism and its corresponding production cross section has been found to range from $85$ fb up to $10$ fb for $Z^\prime $ gauge boson masses between $4$ TeV and $5$ TeV and LHC center of mass energy $\sqrt{S}=13$ TeV \cite{Long:2018dun}. Such $Z^\prime$ gauge boson after being produced will decay into pair of SM particles, with dominant decay mode into quark-antiquark pairs as shown in detail in Refs. \cite{Perez:2004jc,CarcamoHernandez:2005ka}. The two body decays of the $Z^\prime$ gauge boson in 3-3-1 models have been studied in details in Refs. \cite{Perez:2004jc}. In particular, in Ref. \cite{Perez:2004jc} it has been shown the $Z^\prime$ decays into a lepton pair in 3-3-1 models have branching ratios of the order of $10^{-2}$, which implies that the total LHC cross section for the $pp\to Z^\prime\to l^{+}l^{-}$ resonant production at $\protect\sqrt{S}=13$ TeV will be of the order of $1$ fb for a $4$ TeV $Z^\prime$ gauge boson, which is below its corresponding lower experimental limit arising from LHC searches \cite{Aaboud:2017sjh}. A detailed study of the collider phenomenology of this model is beyond the scope of this paper and is left for future studies.

\newpage

\section{Quark masses and mixings}
\label{quarkmassesandmixing}
In this section, we show that our model is able to reproduce the observed pattern of SM quark masses and mixings. From the quark Yukawa
terms, it follows that the up-type mass matrix in the basis $(\overline{u%
}_{1L},\overline{u}_{2L},\overline{u}_{3L},\overline{T}_{L},\overline{%
\widetilde{T}}_{1L},\overline{\widetilde{T}}_{3L})$ versus $%
(u_{1R},u_{2R},u_{3R},T_{R},\widetilde{T}_{1R},\widetilde{T}_{2R})$ takes the form: 
\begin{eqnarray}
M_{U} &=&\left( 
\begin{array}{cccc}
0_{2\times 2} & 0_{2\times 1} & 0_{2\times 1} & A_{U} \\ 
0_{1\times 2} & m_{t} & 0 & 0_{1\times 2} \\ 
0_{1\times 2} & 0 & M_{T} & 0_{1\times 2} \\ 
B_{U} & 0_{2\times 1} & 0_{2\times 1} & M_{\widetilde{T}}%
\end{array}%
\right) ,\hspace{1cm}A_{U}=y^{\left( U\right) }\frac{vv_{\chi }}{2\Lambda }%
\left( 
\begin{array}{cc}
0 & 1 \\ 
-1 & 0%
\end{array}%
\right) ,\hspace{1cm}  \notag \\
B_{U} &=&\left( 
\begin{array}{cc}
x_{11}^{\left( U\right) }\left( \frac{v_{\sigma }}{\Lambda }\right)
^{5}v_{S_{1}} & x_{12}^{\left( U\right) }\frac{v_{S_{6}}v_{\sigma }}{\Lambda
^{2}}v_{S_{1}} \\ 
0 & x_{22}^{\left( U\right) }\frac{v_{S_{5}}}{\Lambda }v_{S_{2}}%
\end{array}%
\right) =\left( 
\begin{array}{cc}
z_{11}^{\left( U\right) }\lambda ^{4} & z_{12}^{\left( U\right) }\lambda \\ 
0 & z_{22}^{\left( U\right) }%
\end{array}%
\right) \lambda v_{S_{1}},\hspace{1cm}  \notag \\
M_{T} &=&m_{\widetilde{T}}\left( 
\begin{array}{cc}
0 & 1 \\ 
-1 & 0%
\end{array}%
\right) ,\hspace{1cm}m_{t}=y_{3}^{\left( U\right) }\frac{v}{\sqrt{2}},%
\hspace{1cm}m_{T}=y^{\left( T\right) }\frac{v_{\chi }}{\sqrt{2}},  \label{MU}
\end{eqnarray}
while the down type quark mass matrix written in the basis\newline 
$(\overline{d}_{1L},\overline{d}_{2L},\overline{d}_{3L},\overline{J}_{1L},\overline{J}_{2L},\overline{B}_{1L},\overline{B}_{2L},\overline{B}_{3L})$-$(d_{1R},d_{2R},d_{3R},J_{1R},J_{2R},B_{1R},B_{2R},B_{3R})$ reads: 
\begin{eqnarray}
M_{D} &=&\left( 
\begin{array}{ccc}
0_{3\times 3} & 0_{3\times 2} & A_{D} \\ 
0_{2\times 3} & M_{J} & 0_{2\times 3} \\ 
B_{D} & 0_{3\times 2} & M_{B}%
\end{array}%
\right) ,\hspace{1cm}  \notag \\
A_{D} &=&\left( 
\begin{array}{ccc}
y_{11}^{\left( D\right) }\frac{v_{S_{1}}v_{\sigma }^{2}}{\Lambda ^{3}} & 0 & 
y_{13}^{\left( D\right) }\frac{v_{S_{3}}v_{\sigma }}{\Lambda } \\ 
0 & y_{22}^{\left( D\right) }\frac{v_{S_{2}}}{\Lambda } & y_{23}^{\left(
D\right) }\frac{v_{S_{2}}}{\Lambda } \\ 
0 & 0 & y_{33}^{\left( D\right) }\frac{v_{\chi }}{\Lambda }%
\end{array}%
\right) \frac{v}{\sqrt{2}}=\left( 
\begin{array}{ccc}
z_{11}^{\left( D\right) }\lambda ^{5} & 0 & z_{13}^{\left( D\right) }\lambda
^{4} \\ 
0 & z_{22}^{\left( D\right) }\lambda ^{3} & z_{23}^{\left( D\right) }\lambda
^{3} \\ 
0 & 0 & z_{3}^{\left( D\right) }\lambda%
\end{array}%
\right) \frac{v}{\sqrt{2}},  \notag \\
B_{D} &=&\left( \frac{v_{\sigma }}{\Lambda }\right) ^{2}\left( 
\begin{array}{ccc}
x_{1}^{\left( D\right) } & 0 & 0 \\ 
0 & x_{2}^{\left( D\right) } & 0 \\ 
0 & 0 & x_{3}^{\left( D\right) }%
\end{array}%
\right) v_{S_{4}}=\left( 
\begin{array}{ccc}
z_{1}^{\left( D\right) } & 0 & 0 \\ 
0 & z_{2}^{\left( D\right) } & 0 \\ 
0 & 0 & z_{3}^{\left( D\right) }%
\end{array}%
\right) \lambda ^{2}v_{S_{4}},\hspace{1cm}  \notag \\
M_{J} &=&y^{\left( J\right) }\frac{v_{\chi }}{\sqrt{2}}\left( 
\begin{array}{cc}
0 & 1 \\ 
-1 & 0%
\end{array}%
\right) ,\hspace{1cm}M_{B}=\left( 
\begin{array}{ccc}
m_{B_{1}} & 0 & 0 \\ 
0 & m_{B_{2}} & 0 \\ 
0 & 0 & m_{B_{3}}%
\end{array}%
\right).
\end{eqnarray}
Assuming that the exotic quarks have TeV scale masses, we find that the SM quarks (excepting the top quark) get their
masses from a Universal seesaw mechanism mediated by the two exotic
up-type and three exotic down-type quarks, $\widetilde{T}_{n}$ ($n=1,2$) and $B_{i}$ ($%
i=1,2,3$), respectively. Due to the symmetries of the model, there are no mixing mass terms between the top quark and the remaining up-type quarks. Thus, the Universal Seesaw mechanism gives rise to the following SM quark mass matrices: 
\begin{equation}
\widetilde{M}_{U}=\left( 
\begin{array}{cc}
A_{U}M_{\widetilde{T}}^{-1}B_{U} & 0_{2\times 1} \\ 
0_{1\times 2} & m_{t}%
\end{array}%
\right) =\left( 
\begin{array}{ccc}
z_{11}^{\left( U\right) }y^{\left( U\right) }\frac{\lambda
^{5}vv_{S_{1}}v_{\chi }}{2\Lambda m_{\widetilde{T}}} & z_{12}^{\left(
U\right) }y^{\left( U\right) }\frac{\lambda ^{2}vv_{S_{1}}v_{\chi }}{%
2\Lambda m_{\widetilde{T}}} & 0 \\ 
0 & z_{22}^{\left( U\right) }y^{\left( U\right) }\frac{\lambda
vv_{S_{1}}v_{\chi }}{2\Lambda m_{\widetilde{T}}} & 0 \\ 
0 & 0 & m_{t}%
\end{array}%
\right) =\left( 
\begin{array}{ccc}
a_{11}\lambda ^{8} & a_{12}\lambda ^{5} & 0 \\ 
0 & a_{22}\lambda ^{4} & 0 \\ 
0 & 0 & \alpha%
\end{array}%
\right) \allowbreak \allowbreak \frac{v}{\sqrt{2}},
\end{equation}%
\begin{equation}
\widetilde{M}_{D}=A_{D}M_{B}^{-1}B_{D}=\left( 
\begin{array}{ccc}
z_{11}^{\left( D\right) }\lambda ^{5} & 0 & z_{13}^{\left( D\right) }\lambda
^{4} \\ 
0 & z_{22}^{\left( D\right) }\lambda ^{3} & z_{23}^{\left( D\right) }\lambda
^{3} \\ 
0 & 0 & z_{3}^{\left( D\right) }\lambda%
\end{array}%
\right) \frac{\lambda ^{2}v_{S_{4}}v}{\sqrt{2}m_{B}}=\left( 
\begin{array}{ccc}
b_{11}\lambda ^{7} & 0 & b_{13}\lambda ^{6} \\ 
0 & b_{22}\lambda ^{5} & b_{23}\lambda ^{5} \\ 
0 & 0 & b_{33}\lambda ^{3}%
\end{array}%
\right) \allowbreak \frac{v}{\sqrt{2}},
\end{equation}%
where we have set $\left( m_{B}\right) _{j}=m_{B}$ ($j=1,2,3$) and
considered $m_{\widetilde{T}}\sim m_{B}\sim v_{\chi }\sim v_{S_{4}}$. Let us
note that in our model, the dominant contribution to the Cabbibo mixing
arises from the up-type quark sector, whereas the down-type quark sector
contributes to the remaining CKM mixing angles. Given that we are considering real Yukawa couplings, in order to account for CP
violation in the quark sector we take $v_{S_{3}}$ to be complex, which
implies that the only complex entry in the SM quark mass matrices is $b_{13}$. Thus, in this scenario, and taking into account that the scalar $S_3$ is a $T^{\prime}$ doublet charged under the $Z_{12}$ symmetry as shown in Table \ref{tab:scalars}, the observed CP violation in the quark sector will arise from the spontaneous breaking of the $T^{\prime}\times Z_{12}$ discrete group by the vacuum expectation value of the $S_3$ scalar.

The experimental values of the physical quark mass spectrum \cite%
{Bora:2012tx,Xing:2007fb}, mixing angles and Jarlskog invariant \cite%
{Olive:2016xmw} can be obtained from the following benchmark point: 
\begin{eqnarray}
a_{11} &\simeq &1.259,\hspace{1cm}a_{12}\simeq -1.441,\hspace{1cm}%
a_{22}\simeq 1.400,\hspace{1cm}\alpha \simeq 0.989,\hspace{1cm}b_{11}\simeq
0.579,  \notag  \label{eq:Quark-benchmark-point} \\
b_{22} &\simeq &0.604,\hspace{1cm}\left\vert b_{13}\right\vert \simeq 1.265,%
\hspace{1cm}\arg \left( b_{13}\right) \simeq -158^{\circ },\hspace{1cm}%
b_{23}\simeq 1.117,\hspace{1cm}b_{33}\simeq 1.431.
\end{eqnarray}%
\begin{table}[tbh]
\begin{center}
\begin{tabular}{c|l|l}
\hline\hline
Observable & Model value & Experimental value \\ \hline
$m_{u}(\mathrm{MeV})$ & \quad $1.38$ & \quad $1.45_{-0.45}^{+0.56}$ \\ \hline
$m_{c}(\mathrm{MeV})$ & \quad $635$ & \quad $635\pm 86$ \\ \hline
$m_{t}(\mathrm{GeV})$ & \quad $172.1$ & \quad $172.1\pm 0.6\pm 0.9$ \\ \hline
$m_{d}(\mathrm{MeV})$ & \quad $2.9$ & \quad $2.9_{-0.4}^{+0.5}$ \\ \hline
$m_{s}(\mathrm{MeV})$ & \quad $60.0$ & \quad $57.7_{-15.7}^{+16.8}$ \\ \hline
$m_{b}(\mathrm{GeV})$ & \quad $2.82$ & \quad $2.82_{-0.04}^{+0.09}$ \\ \hline
$\sin \theta _{12}$ & \quad $0.225$ & \quad $0.225$ \\ \hline
$\sin \theta _{23}$ & \quad $0.0412$ & \quad $0.0412$ \\ \hline
$\sin \theta _{13}$ & \quad $0.00365$ & \quad $0.00365$ \\ \hline
$J$ & \quad $3.30\times 10^{-5}$ & \quad $\left( 3.18\pm 0.15\right) \times
10^{-5}$ \\ \hline\hline
\end{tabular}%
\end{center}
\caption{Model and experimental values of the quark masses and CKM
parameters.}
\label{Tab:quarks}
\end{table}\newline

As indicated in Table \ref{Tab:quarks} our model successfully reproduces the low energy quark flavor data by having the quark model parameters of order unity. The symmetries of our model give rise to quark mass matrix textures that successfully explain the SM quark mass spectrum and mixing parameters, without requiring the introduction of a hierarchy in the free effective parameters of the quark sector. These effective parameters only need to be mildly tuned in order to perfectly reproduce the observed quark mass spectrum and CKM parameters.

Finaly to close this section we briefly comment about the LHC signatures of exotic quarks in our model. As follows from the quark Yukawa terms of Eq. (\ref{eqn:Lyq}), the exotic quarks have mixing mass terms with all SM quarks, excepting the top quark. Such mixing terms allow that these exotic quarks can decay into any of the scalars of the model and a SM quark. These exotic quarks can decay into a SM quark and the SM-like Higgs boson. Such exotic quarks can be produced in pairs at the LHC via gluon fusion and Drell–Yan mechanism. Consequently, observing an excess of events in the six jet final state can be a signal of support of this model at the LHC. A detailed study of the exotic quark production at the LHC and the exotic quark decay modes is beyond the scope of this work and is left for future studies.

\section{Lepton masses and mixings.}
\label{leptonsector}
From the charged lepton Yukawa interactions given in Eq. (\ref{eqn:lyl})  and using Eqs. (\ref{eqn:vevt}) and (\ref{VEVsinglets}) together with the product rules of the $T^{\prime }$ group shown in the Appendix, we find that the
charged lepton matrix is given by: 
\begin{equation}
M_{l}=%
\begin{pmatrix}
e^{-i\alpha }\cos \gamma  & -e^{-i\alpha }\sin \gamma  & 0 \\ 
e^{i\alpha }\sin \gamma  & e^{i\alpha }\cos \gamma  & 0 \\ 
0 & 0 & 1%
\end{pmatrix}%
\begin{pmatrix}
f_{1}\lambda ^{9} & e^{i(\phi _{2}-\phi _{1})}f_{2}\lambda ^{5} & 
e^{-i(2\phi _{2}+\phi _{1})}f_{3}\lambda ^{3} \\ 
e^{i(2\phi _{1}+\phi _{2})}f_{1}\lambda ^{9} & f_{2}\lambda ^{5} & 
e^{-i(2\phi _{2}+\phi _{1})}f_{3}\lambda ^{3} \\ 
e^{i(2\phi _{1}+\phi _{2})}f_{1}\lambda ^{9} & e^{i(\phi _{2}-\phi
_{1})}f_{2}\lambda ^{5} & f_{3}\lambda ^{3}%
\end{pmatrix}%
\dfrac{v}{\sqrt{2}},\label{eqn:ml}
\end{equation}
where $f_i$ with $i=1,2,3$ are $\mathcal{O}(1)$ dimensionless parameters assumed to be real.

Regarding the neutrino sector, from the lepton Yukawa terms given in Eq. (\ref{eqn:lyl}), we find the following neutrino mass terms: 
\begin{align}
-\mathcal{L}^{(\nu)}_{\text{mass}} &= \dfrac{1}{2} 
\begin{pmatrix}
\overline{\nu^C_L} & \overline{\nu_R} & \overline{N_R}%
\end{pmatrix}
M_\nu 
\begin{pmatrix}
\nu_L \\ 
\nu^C_R \\ 
N^C_R%
\end{pmatrix}
+ H.c,
\label{Lynumass}
\end{align}
where the neutrino mass matrix $M_\nu$ is
\begin{align}
M_\nu &= 
\begin{pmatrix}
0_{3\times3} & M_{\nu_D} & 0_{3\times3} \\ 
M_{\nu_D}^T & 0_{3\times3} & M_\chi \\ 
0_{3\times3} & M_\chi^T & M_R%
\end{pmatrix},%
\end{align}
and the submatrices $M_{\nu_D}$ and $M_\chi$ are generated from the $y_{\rho}\varepsilon_{abc}\varepsilon_{dec} \left(\overline{L}_{L}^{a}\left( L_{L}^{C}\right)^{b}\right)_{\mathbf{3}_{2}} \eta^{d}
\chi^{e} \frac{\xi\sigma^{2}}{\Lambda^{4}}$ and $y_\chi^{(L)}\left(\overline{L}_{L} \chi N_{R}\right)_{\mathbf{1}}$ Yukawa terms in Eq. \eqref{eqn:lyl}, respectively. Furthermore, the submatrix $M_R$ arises from the Majorana neutrino Yukawa interactions shown in the third line of Eq. \eqref{eqn:lyl}. The submatrices $M_{\nu_D}$, $M_\chi$ and $M_R$ take the form: 
\begin{eqnarray}
M_{\nu _{D}} &=&%
\begin{pmatrix}
0 & Ae^{-i(2\phi_{1}+\phi_{2})} & -Ae^{-i(2\phi_{1}+\phi_{2})} \\ 
-Ae^{-i(2\phi_{1}+\phi_{2})} & 0 & A \\ 
Ae^{-i(2\phi_{1}+\phi_{2})} & -A & 0%
\end{pmatrix},  \label{eqn:md}
\\
M_{\chi } &=&%
\begin{pmatrix}
B & 0 & 0 \\ 
0 & 0 & Be^{i(2\phi _{1}+\phi_{2})} \\ 
0 & Be^{i(2\phi _{1}+\phi_{2})} & 0
\end{pmatrix},  \label{eqn:mx} \\
M_{R} &=&%
\begin{pmatrix}
(C+2D) & -D e^{i(2\phi_{1}+\phi_{2})} & -D e^{i(2\phi_{1}+\phi_{2})} \\ 
-D e^{i(2\phi _{1}+\phi _{2})} & 2De^{3i\phi _{1}} & (C-D)e^{i(2\phi_{1}+\phi_{2})} \\ 
-D e^{i(2\phi_{1}+\phi_{2})} & (C-D)e^{i(2\phi_{1}+\phi_{2})} & 2D e^{3i(\phi_{1}+\phi_{2})} 
\end{pmatrix}. \label{eqn:mr}
\end{eqnarray}
where $A$, $B$, $C$ and $D$ are given by:
\begin{equation}
A=\frac{y_{\rho }v_{\eta }v_{\chi }v_{\xi }v_{\sigma }^{2}}{2\sqrt{2}\Lambda
^{4}},\hspace{1.5cm}B=\frac{y_{\chi }^{\left( L\right) }v_{\chi }}{%
\sqrt{2}},\hspace{1.5cm}C=h_{1N}\frac{v_{\sigma }^{6}v^{4}_{S_6}}{%
\Lambda ^{10}}v_{\varphi },\hspace{1.5cm}D=h_{2N}\frac{v_{\xi }v_{\sigma }^{6}v^{4}_{S_6}}{%
\Lambda ^{11}}v_{\varphi }.
\label{ParametersABCD}
\end{equation}
As shown in detail in Ref. \cite{Catano:2012kw}, the full rotation matrix that diagonalizes the neutrino mass matrix $M_\nu$ is approximately given by
\begin{align}
\mathbb{U} &= 
\begin{pmatrix}
V_\nu & X_3 U_\chi & X_2 U_R \\
-\frac{(X_2^\dagger + X_3^\dagger)}{\sqrt{2}}V_\nu & \frac{(1-S)}{\sqrt{2}}U_\chi & \frac{(1+S)}{\sqrt{2}} U_R \\
-\frac{(X_2^\dagger - X_3^\dagger)}{\sqrt{2}}V_\nu & \frac{(-1-S)}{\sqrt{2}}U_\chi & \frac{(1-S)}{\sqrt{2}}U_R 
\end{pmatrix},
\end{align}
where 
\begin{align}
S &= -\dfrac{1}{2 \sqrt{2}\,y_\chi^{(L)} v_\chi} M_R, & X_2 &\simeq X_3 \simeq	\dfrac{1}{y_\chi^{(L)} v_\chi} M_{\nu_{D}}^*,
\end{align}
and the physical neutrino mass matrices are:
\begin{align}
M_\nu^{(1)} &= M_{\nu_D} (M_\chi^T)^{-1} M_R\,M_\chi^{-1} M_{\nu_D}^T,  \label{eqn:seesaw}
\end{align}
\begin{align}
M_\nu^{(2)} &= -\dfrac{1}{2} (M_\chi + M_\chi^T) + \dfrac{1}{2} M_R, &
M_\nu^{(3)} &= \dfrac{1}{2} (M_\chi + M_\chi^T) + \dfrac{1}{2} M_R, 
\end{align}
where $M_\nu^{(1)}$ is the light active neutrino mass matrix whereas $M_\nu^{(2)}$ and $M_\nu^{(3)}$ are the exotic Dirac neutrino mass matrices. The physical neutrino spectrum is composed of 3 light active neutrinos and 6 nearly degenerate sterile exotic pseudo-Dirac neutrinos.

Furthermore, from Eqs. (\ref{eqn:md})-(\ref{ParametersABCD}) and (\ref{eqn:seesaw}), we find for the light active neutrino mass scale, the estimate $m_{\nu}\sim\lambda^{22}v_{\varphi}\sim 50$ meV. Consequently, our model provides a natural explanation for the smallness of the light active neutrino masses. \newline

The sterile neutrinos can be pair produced at the Large Hadron Collider (LHC), via a Drell-Yan annihilation mediated by a heavy $Z^\prime $ gauge boson. The sterile neutrinos mix the light active ones thus allowing the sterile neutrinos to decay into SM particles, so that the final decay products will be a SM charged lepton and a $W$ gauge boson. Consequently, the observation of an excess of events in the dilepton final states above the SM background, can be a signal in support of this model at the LHC. Studies of inverse seesaw neutrino signatures at the colliders as well as the production of heavy neutrinos at the LHC are carried out in Refs. \cite{BhupalDev:2012zg,Das:2012ze,Das:2014jxa,Das:2016hof,Das:2017gke,Das:2017nvm,Das:2017zjc,Das:2017rsu,Das:2018usr,Das:2018hph,Bhardwaj:2018lma,Helo:2018rll,Pascoli:2018heg}. A comprehensive study of the implications of our model at colliders goes beyond the scope of this work and will be done elsewhere.\newline
 
\begin{figure}[H]
    \begin{subfigure}{.5\linewidth}
        \centering
        \captionsetup{width=0.8\textwidth}
        \includegraphics[scale=.55]{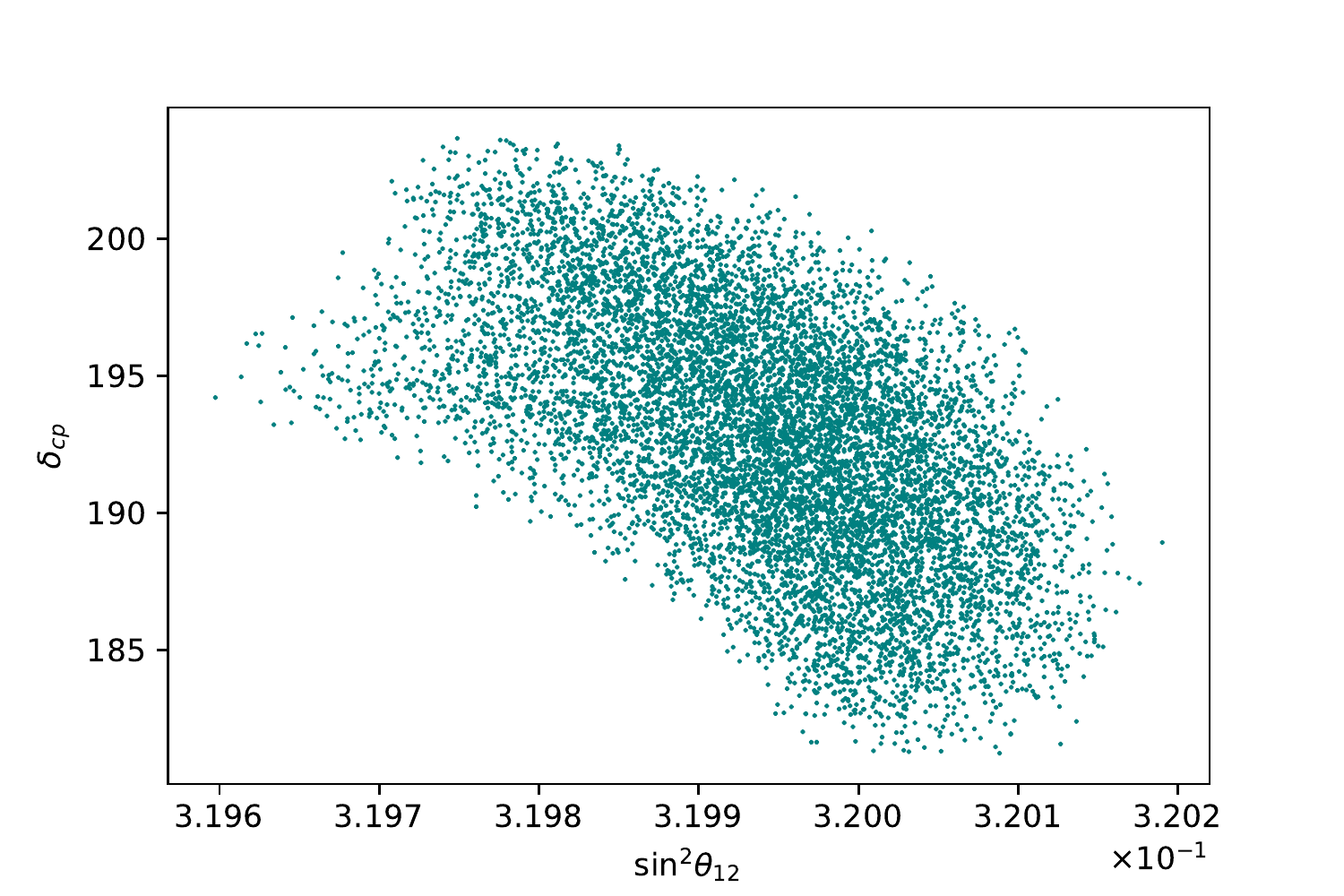}
        \caption{Correlation between the solar mixing parameter $\sin^2 \theta_{12}$ and the leptonic Dirac CP-violating phase $\delta_{CP}$.}
        \label{fig:sub1}
    \end{subfigure}
    \begin{subfigure}{.5\linewidth}
        \centering
        \captionsetup{width=0.8\textwidth}
        \includegraphics[scale=.55]{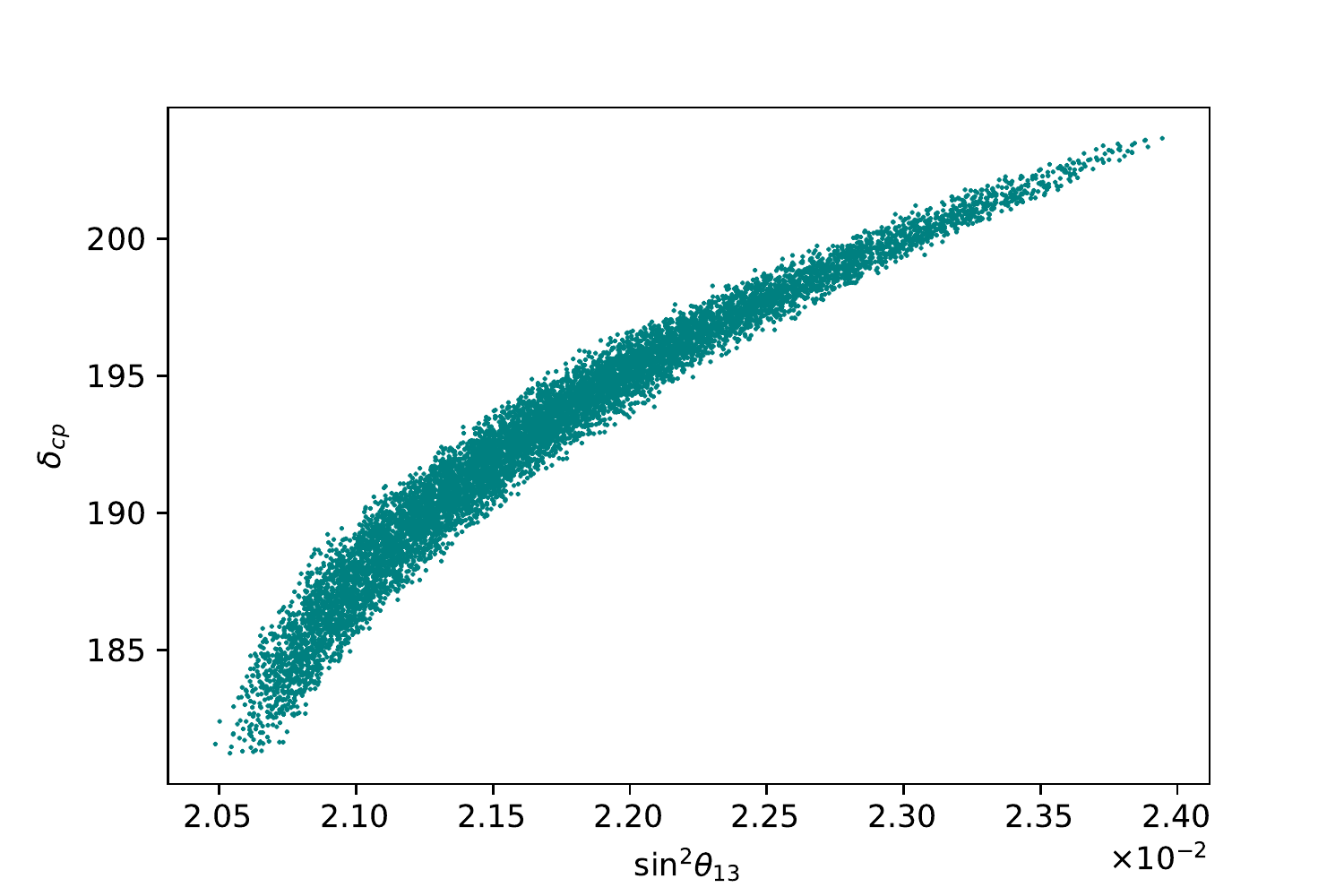}
        \caption{Correlation between the reactor mixing parameter $\sin^2 \theta_{13}$ and the leptonic Dirac CP-violating phase $\delta_{CP}$.}
        \label{fig:sub2}
    \end{subfigure}\\[1ex]
    \begin{subfigure}{\linewidth}
        \centering
        \captionsetup{width=0.43\textwidth}
        \includegraphics[scale=.55]{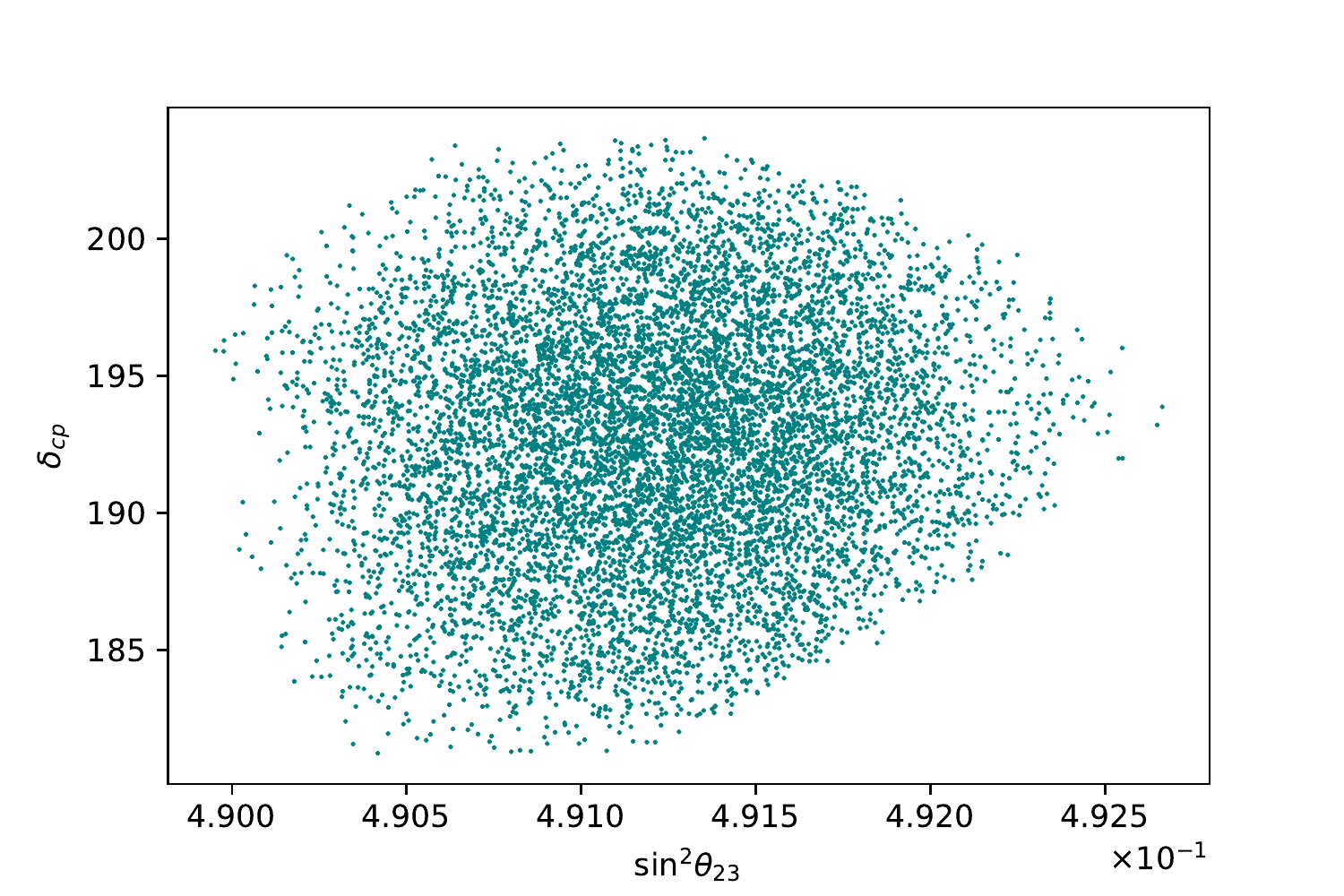}
        \caption{Correlation between the atmospheric mixing parameter $\sin^2 \theta_{23}$ and the leptonic Dirac CP-violating phase $\delta_{CP}$.}
        \label{fig:sub3}
    \end{subfigure}
    \caption{Correlations between the leptonic mixing parameters and the leptonic Dirac CP violating phase.} 
    \label{fig:mixingangledeltacp}
\end{figure}

By varying the lepton sector model parameters, we obtain values for the charged lepton masses, neutrino mass squared differences and leptonic mixing parameters in very good agreement with the experimental data, as shown in Table \ref{Tab:neutrinofit}. This shows that our model can successfully accommodate the experimental values of the physical observables of the lepton sector. It is worth mentioning that the range for the experimental values in Table \eqref{Tab:neutrinofit} were taken from \cite{deSalas:2017kay} for the case of normal hierarchy. Let us note that we only consider the case of normal hierarchy since it is favored over more than $3\sigma$ than the inverted neutrino mass ordering. Furthermore, let us note that given that we are considering real Yukawa couplings in our model, the observed CP violation in the lepton sector is generated by the spontaneous breaking of the $T^{\prime}\times Z_6\times Z_8$ discrete group by the vacuum expectation values of the $\rho$, $\phi$, $\zeta$ and $\xi$ scalars. 
\begin{table}[H]
\vspace{-0.5cm}
\centering
\begin{tabular}{|c|c|c|c|c|}
\hline
\multirow{2}{*}{Observable} & \multirow{2}{*}{Model Value} & 
\multicolumn{3}{c|}{Experimental value} \\ \cline{3-5}
&  & $1\sigma$ range & $2\sigma$ range & $3\sigma$ range \\ \hline\hline
$m_{e}$ [MeV] & $0.487$ & $0.487$ & $0.487$ & $0.487$ \\ \hline
$m_{\mu }$ [MeV] & $102.8$ & $102.8\pm 0.0003$ & $102.8\pm 0.0006$ & $%
102.8\pm 0.0009$ \\ \hline
$m_{\tau }$ [GeV] & $1.75$ & $1.75\pm 0.0003$ & $1.75\pm 0.0006$ & $1.75\pm
0.0009$ \\ \hline
$\Delta m_{21}^2$ $[10^{-5}\,eV^2]$ & $7.54987$ & $7.55^{+0.20}_{-0.16}$ & $7.20 -
7.94$ & $7.05 - 8.14$ \\ \hline
$\Delta m_{31}^2$ $[10^{-3}\,eV^2]$ & $2.49995$ & $2.50\pm0.03$ & $2.44 - 2.57$ & $%
2.41-2.60$ \\ \hline
$\sin^2(\theta_{12})/10^{-1}$ & $3.19999$ & $3.20^{+0.20}_{-0.16}$ & $%
2.89-3.59$ & $2.73-3.79$ \\ \hline
$\sin^2(\theta_{23})/10^{-1}$ & $4.91197$ & $5.47^{+0.20}_{-0.30}$ & $%
4.67-5.83$ & $4.45-5.99$ \\ \hline
$\sin^2(\theta_{13})/10^{-2}$ & $2.16073$ & $2.160^{+0.083}_{-0.069}$ & $%
2.03-2.34$ & $1.96-2.41$ \\ \hline
$\delta_{CP}$ & $192.761^\circ$ & $218^{+38^\circ}_{-27^\circ}$ & $182^\circ-315^\circ$ & $157^\circ-349^\circ$ \\
\hline
\end{tabular}
\caption{The model values shown in the table are the best fit results for the neutrino mass squared differences, mixing angles and the CP-violating phase for the case of normal hierarchy. The 1-3$\sigma$ experimental ranges \protect\cite{deSalas:2017kay} are also shown for comparison.}
\label{Tab:neutrinofit}
\end{table}
  Figure \ref{fig:mixingangledeltacp} shows the correlation between the leptonic mixing parameters and the leptonic Dirac CP violating phase for the case of normal neutrino mass hierarchy. To obtain these Figures, the lepton sector parameters were randomly generated in a range of values where the neutrino mass squared splittings and leptonic mixing parameters are inside the $3\sigma$ experimentally allowed range. We found a leptonic Dirac CP violating phase in the range $180^{\circ}\lesssim\delta_{CP}\lesssim 205^{\circ}$, whereas the leptonic mixing parameters are obtained to be in the ranges $0.3196\lesssim\sin^2\theta_{12}\lesssim 0.3202$, $0.4900\lesssim\sin^2\theta_{23}\lesssim 0.4925$ and $0.0205\lesssim\sin^2\theta_{13}\lesssim 0.0240$. 

\section{Higgs diphoton rate}
\label{htogammagamma}

The explicit form of the $h \rightarrow \gamma\gamma$ decay rate is \cite{Shifman:1979eb,Gavela:1981ri,Kalyniak:1985ct,Djouadi:2005gj,Marciano:2011gm,Wang:2012gm}
\begin{align}
\Gamma(h \rightarrow \gamma\gamma) = \dfrac{\alpha_{em}^2 m_h^3}{256 \pi^3 v^2} \biggr| \sum_f a_{hff} N_C Q_f^2 F_{1/2}(\rho_f) + a_{hWW} F_{1}(\rho_W) + a_{hW'W'} F_{1}(\rho_{W'}) \biggr|^2.
\end{align}
Here $\rho_i$ are the mass ratios $\rho_i= \frac{m_h^2}{4 M_i^2}$ with $M_i=m_f, M_W, M_{W'}$; $\alpha_{em}$ is the fine structure constant; $N_C$ is the color factor ($N_C=1$ for leptons and $N_C=3$ for quarks); and $Q_f$ is the electric charge of the fermion in the loop.

From the fermionic loop contributions, we only consider the one arising from the top quark exchange. Furthermore, $a_{htt}$, $a_{hWW}$ and $a_{hW'W'}$ are the deviation factors from the SM expectation, of the Higgs–top quark coupling, the Higgs–WW and the Higgs–W'W' gauge boson couplings, respectively:
\begin{align}
    a_{hWW} &= -\sin \alpha, \\
    a_{hW'W'} &= \cos \alpha \cot \beta, \\
    a_{htt} &\simeq 1.
    \label{aparameters}
\end{align}
The numerical values of these parameters are given in Table \ref{tab:hgammagamma}. Let us note that in our model the Higgs–top quark coupling is very close to the SM expectation, i.e., $a_{htt}\simeq 1$, since the mixing between the CP even neutral scalar fields $\xi_{\eta}$ and $\xi_{\chi}$ is very suppressed, being the $126$ GeV SM like Higgs boson mainly composed of the $\xi_{\eta}$ field.

The dimensionless loop factors $F_{1/2}(\rho)$ and $F_{1}(\rho)$ for spin-$1/2$ and spin-$1$ particles in the loop, respectively are \cite%
{Shifman:1979eb,Gavela:1981ri,Kalyniak:1985ct,Gunion:1989we,Spira:1997dg,Djouadi:2005gj,Marciano:2011gm,Wang:2012gm}: 
\begin{align}
    F_{1/2}(\rho) &= 2(\rho + (\rho -1)f(\rho))\rho^{-2}, \\
    F_{1}(\rho) &= -2(2\rho^2 + 3\rho + 3(2\rho-1)f(\rho))\rho^{-2}, \\
    F_{0} &= -(\rho - f(\rho))\rho^{-2},
\end{align}
with
\begin{align}
    f(\rho) = \left\{ \begin{array}{lcc}
             \arcsin^2 \sqrt{2} & \text{for}  & \rho \leq 1 \\
             \\ -\frac{1}{4}\left(\ln \left(\frac{1+\sqrt{1-\rho^{-1}}}{1-\sqrt{1-\rho^{-1}}-i\pi} \right)^2 \right) &  \text{for} & \rho > 1.\\
             \end{array}
   \right.
\end{align}
In what follows we show that our model is consistent with the current Higgs diphoton decay rate constraints. 
To this end, we introduce the ratio $R_{\gamma \gamma}$, which normalizes the $\gamma \gamma$ signal predicted by our model relative to that of the SM:
\begin{align}
    R_{\gamma \gamma} = \frac{\sigma(pp \to h)\Gamma(h \to \gamma \gamma)}{\sigma(pp \to h)_{SM}\Gamma(h \to \gamma \gamma_{SM}} \simeq a^{2}_{htt} \frac{\Gamma(h \to \gamma \gamma)}{\Gamma(h \to \gamma \gamma)_{SM}}.
    \label{eqn:hgg}
\end{align}
The normalization given by \eqref{eqn:hgg} for $h \to \gamma \gamma$ was also used in \cite{Campos:2014zaa,Wang:2012gm,Carcamo-Hernandez:2013ypa,Bhattacharyya:2014oka,Fortes:2014dia,Hernandez:2015xka,Hernandez:2015nga}.

The ratio $R_{\gamma \gamma}$ has been measured by CMS and ATLAS with the best fit signals \cite{Khachatryan:2014ira,Aad:2014eha}:
\begin{align}
    R^{CMS}_{\gamma \gamma} = 1.14^{+0.26}_{-0.23} \ \text{and} \ R^{ATLAS}_{\gamma \gamma} = 1.17 \pm 0.27.
\end{align}

With the best fit results shown in Table \ref{tab:hgammagamma} the $R_{\gamma \gamma}$ parameter has been calculated as:
\begin{equation}
    R_{\gamma \gamma} = 1.0021.
\end{equation} 
Consequently, our model successfully accommodates the current Higgs diphoton decay rate constraints.
\begin{table}[H]
\centering
\begin{tabular}{c|c}
\hline\
\textbf{Parameters} & \textbf{Model value}  \\
\hline\hline
$a_{hW^-W^+}$ & $0.999816$ \\
\hline
$a_{hW'W'}$ & $0.000471445$  \\
\hline
$a_{htt}$ & $1.0$  \\
\hline
\end{tabular}
\caption{Numerical values for the deviation factors $a_{htt}$, $a_{hWW}$ and $a_{hW'W'}$ used for the computation of the Higgs diphoton decay rate. Here we set $v_\chi=10$ TeV.}
\label{tab:hgammagamma}
\end{table}
\section{Lepton flavor violating constraints}
\label{LFV} 
In this section we will determine the constraints on the model
parameter space imposed by the charged lepton flavor violating processes $%
\mu \rightarrow e\gamma $, $\tau \rightarrow \mu \gamma $ and $\tau
\rightarrow e\gamma $. As mentioned in the previous section, the sterile
neutrino spectrum of the model is composed of six nearly degerate heavy
neutrinos. These sterile neutrinos together with the heavy $W^{\prime }$
gauge boson induce the $l_{i}\rightarrow l_{j}\gamma $ decay at one loop
level, whose Branching ratio is given by: \cite%
{Ilakovac:1994kj,Deppisch:2004fa,Lindner:2016bgg}: 
\begin{equation}
Br\left( l_{i}\rightarrow l_{j}\gamma \right) =\frac{\alpha
_{W}^{3}s_{W}^{2}m_{l_{i}}^{5}}{256\pi ^{2}m_{W^{\prime }}^{4}\Gamma _{i}}%
\left\vert G\left( \frac{m_{N}^{2}}{m_{W^{\prime }}^{2}}\right) \right\vert
^{2},\hspace{0.5cm}\hspace{0.5cm}\hspace{0.5cm}G\left( x\right) =-\frac{%
2x^{3}+5x^{2}-x}{4\left( 1-x\right) ^{2}}-\frac{3x^{3}}{2\left( 1-x\right)
^{4}}\ln x.
\end{equation}
where the one loop level contribution arising from the $W$ gauge boson exchange has been neglected because it is suppressed by the quartic power of the active-sterile neutrino mixing angle $\theta$, which in our model is of the order of $10^{-3}$, for sterile neutrino masses of about $1$ TeV. It has been shown in Ref. \cite{Deppisch:2013cya}, that for such mixing angle the contribution of the $W$ gauge boson to the branching ratio for the $\mu\to e\gamma$ decay rate takes values of the order of $10^{-16}$, which corresponds to three orders of magnitude below its experimental upper limit of $4.2\times 10^{-13}$. Thus, in this work, we only consider the dominant $W^{\prime}$ contribution to the $\mu\to e\gamma$ decay rate.

\begin{figure}[tbp]
\includegraphics[width=0.49\textwidth]{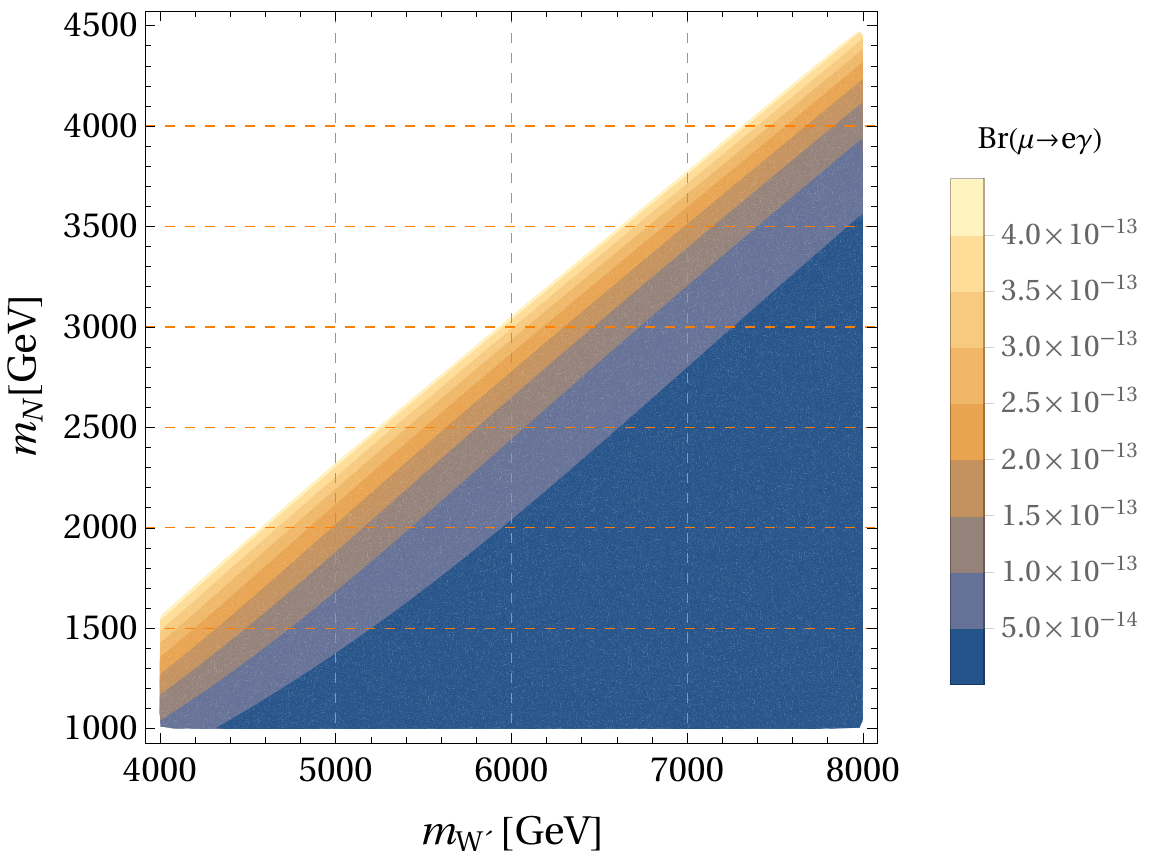}
\caption{Allowed parameter space in the $m_{W^{\prime }}-m_{N}$ plane
consistent with the LFV constraints.}
\label{LFVplot}
\end{figure}

Figure \ref{LFVplot} shows the allowed parameter space in the $m_{W^{\prime
}}-m_{N}$ plane consistent with the constraints arising from charged lepton
flavor violating decays. The $W^{\prime }$ gauge boson and the sterile
neutrino masses have been taken to be in the ranges $4$ TeV$\lesssim
m_{W^{\prime }}\lesssim 8$ TeV and $1$ TeV$\lesssim m_{N}\lesssim
4.5$ TeV, respectively. Notice that we have considered $W^{\prime }$ gauge
boson masses larger than $4$ TeV to fulfill the constraints arising from on $%
K$, $D$ and $B$ meson mixings \cite{Huyen:2012uk}. As seen from Figure \ref%
{LFVplot}, the obtained values for the branching ratio of $\mu \rightarrow
e\gamma $ decay are below its experimental upper limit of $4.2\times
10^{-13} $ and are within the reach of future experimental
sensitivity, in the allowed model parameter space. 
In the region of
parameter space consistent with $\mu\to e\gamma$ decay rate constraints, the maximum obtained branching ratios for the $\tau \rightarrow \mu \gamma $ and $\tau \rightarrow e\gamma $ decays can reach values of the order of $10^{-13}$, which is four orders of magnitude below their corresponding upper experimental bounds of $4.4\times 10^{-8}$ and $3.3\times 10^{-8}$, respectively.  Consequently, our model is compatible with the charged lepton flavor violating decay constaints provided that the sterile neutrino are lighter than about $1.6$ TeV and $4.5$ TeV for $W^{\prime }$ gauge boson masses of $4$ TeV and $8$ TeV, respectively. 
Finally, to close this section we provide some comments regarding the decays of quasi-Dirac sterile neutrinos as well as a comparison of our predictions for such decays and for the LFV signals with the ones obtained in other models with extended gauge symmetry. Notice that, as in the $U^{\prime}(1)$ model of Ref. \cite{Deppisch:2013cya}, in our model the sterile neutrinos feature the two body decay modes: $N\rightarrow l^{\pm}_iW^{\mp}$, $\nu_iZ$ and $\nu_ih$ (where $i=1,2,3$ is a flavor index), which are suppressed by the small active-sterile neutrino mixing angle $\theta\sim\mathcal{O}(10^{-3})$. Those two body decay modes, give rise to the three body decay modes for the sterile neutrinos: $N\rightarrow l^{+}_il^{-}_j\nu_k$, $N\rightarrow l^{-}_iu_j\bar{d}_k$, $N\rightarrow b\bar{b}\nu_k$ (where $i,j,k=1,2,3$ are flavor indices). Such aforementioned decay modes for the sterile neutrinos are also presented in the $U^{\prime}(1)$ model of Ref. \cite{Deppisch:2013cya}. Consequently, we expect similar predictions for the total cross section of the LFV signal process $pp\rightarrow NN\rightarrow e^{\pm}\mu^{\mp}4j$ as well as for the sterile neutrino decays, to the ones obtained in Ref. \cite{Deppisch:2013cya}. A slightly different prediction is expected for the $N\rightarrow l^{+}_il^{-}_j\nu_k$ decay rate, which in our model receives contributions from the diagrams involving the off-shell $W$ and $W^{\prime}$ gauge bosons, whereas in the model of Ref. \cite{Deppisch:2013cya}, it only receives contribution from the diagram involving the off-shell $W$ gauge boson. It is worth mentioning that the contribution to the $N\rightarrow l^{+}_il^{-}_j\nu_k$ decay rate arising from the the exchange of the off-shell $W^{\prime}$ gauge bosons is strongly suppressed by a factor of about $\frac{M^4_W}{M^4_{W^{\prime}}}$ when compared with the one due to the $W$ exchange. Despite such similar expectations for the aforementioned LFV signals, it is worth mentioning that for an active-sterile neutrino mixing angle $\theta\sim\mathcal{O}(10^{-3})$, our obtained values for the branching ratio of the $\mu\to e\gamma$ decay rate will be about three orders of magnitude larger than the obtained in Ref. \cite{Deppisch:2013cya}. This is due to the fact that in our model the $\mu\to e\gamma$ decay is dominated by the $W^{\prime }$ contribution (absent in the $U^{\prime}(1)$ model of Ref. \cite{Deppisch:2013cya}), which is much larger than the contribution arising from the $W$ exchange. Furthermore, we expect similar predictions for the sterile neutrino decay rates and for the cross section of the LFV signal process $pp\rightarrow NN\rightarrow e^{\pm}\mu^{\mp}$ to the ones obtained in the left-right symmetric model of Refs. \cite{AguilarSaavedra:2012fu,Das:2012ii}. Despite such similar predictions, the charged lepton flavor violating process $\mu\to e\gamma$ can be used to discriminate our model from the left-right symmetric model of Refs. \cite{AguilarSaavedra:2012fu,Das:2012ii}, where in the former its dominant contribution arises from the loop diagram involving the $W^{\prime }$ gauge boson exchange, whereas in the latter it receives contributions from the exchange of the $W_R$ gauge boson and the doubly charged scalars contained in the $SU(2)_L$ and $SU(2)_R$ scalar triplets.

\section{Conclusions}
\label{conclusions}
We have constructed a viable 3-3-1 model with two $SU(3)_L$ scalar triplets, extended fermion and scalar
spectrum, based on the $T^{\prime}$ family symmetry and other auxiliary cyclic symmetries, whose spontaneous breaking produces the observed pattern of SM fermion masses and mixing angles. In our model the SM quarks lighter than the top quark, get their masses from a low scale Universal seesaw mechanism, whereas the SM charged lepton masses are produced by a Froggatt-Nielsen mechanism. In addition, the small light active neutrino masses are generated from an inverse seesaw mechanism. Our model is consistent with the low energy SM fermion flavor data and successfully accommodates the current Higgs diphoton decay rate constraints as well as the constraints arising from charged lepton flavor violating processes. In particular, we have found that the constraint on the charged lepton flavor violating decay $\mu\rightarrow e\gamma$ sets the sterile neutrino masses to be lighter than about $1.6$ TeV and $4.5$ TeV for $W^{\prime }$ gauge boson masses of $4$ TeV and $8$ TeV, respectively. We have found that in the allowed region of parameter space, the obtained maximum values of the $\mu\to e\gamma$ branching ratio are close to about $4\times 10^{-13}$, which is within the reach of future experimental sensitivity. Furthermore, the obtained branching ratios for the $\tau\rightarrow \mu \gamma $ and $\tau \rightarrow e\gamma $ decays can reach values of the order of $10^{-13}$. Consequently, our model predicts charged lepton flavor violating decays within the reach of future experimental sensitivity.

\section*{Acknowledgments}

This research has received funding from Fondecyt (Chile), Grants No.~1170803, CONICYT PIA/Basal FB0821, and Programa de Incentivos a la Iniciación Científica (PIIC) from UTFSM (Chile).

\appendix

\section{The product rules for T'}
\label{Tprime}
\def\2tvec#1#2{
\left(
\begin{array}{c}
#1  \\
#2  \\   
\end{array}
\right)}

\def\mat2#1#2#3#4{
\left(
\begin{array}{cc}
#1 & #2 \\
#3 & #4 \\
\end{array}
\right)
}

\def\Mat3#1#2#3#4#5#6#7#8#9{
\left(
\begin{array}{ccc}
#1 & #2 & #3 \\
#4 & #5 & #6 \\
#7 & #8 & #9 \\
\end{array}
\right)
}

\def\3tvec#1#2#3{
\left(
\begin{array}{c}
#1  \\
#2  \\   
#3  \\
\end{array}
\right)}

\def\4tvec#1#2#3#4{
\left(
\begin{array}{c}
#1  \\
#2  \\   
#3  \\
#4  \\
\end{array}
\right)}

\def\5tvec#1#2#3#4#5{
\left(
\begin{array}{c}
#1  \\
#2  \\
#3  \\
#4  \\
#5  \\
\end{array}
\right)}

\def\L{\left}
\def\R{\right}

\def\pl{\partial}

\def\lra{\leftrightarrow}

\def\2tvec#1#2{
\left(
\begin{array}{c}
#1  \\
#2  \\   
\end{array}
\right)}

\def\mat2#1#2#3#4{
\left(
\begin{array}{cc}
#1 & #2 \\
#3 & #4 \\
\end{array}
\right)
}

\def\Mat3#1#2#3#4#5#6#7#8#9{
\left(
\begin{array}{ccc}
#1 & #2 & #3 \\
#4 & #5 & #6 \\
#7 & #8 & #9 \\
\end{array}
\right)
}

\def\3tvec#1#2#3{
\left(
\begin{array}{c}
#1  \\
#2  \\   
#3  \\
\end{array}
\right)}

\def\4tvec#1#2#3#4{
\left(
\begin{array}{c}
#1  \\
#2  \\   
#3  \\
#4  \\
\end{array}
\right)}

\def\5tvec#1#2#3#4#5{
\left(
\begin{array}{c}
#1  \\
#2  \\
#3  \\
#4  \\
#5  \\
\end{array}
\right)}
\def\L{\left}
\def\R{\right}

\def\pl{\partial}

\def\lra{\leftrightarrow}

The double tetrahedral group $T^{\prime }$ is the smallest discrete subgroup of $SU(2)$ as well as the smallest group of any kind with 1-, 2- and 3-dimensional representations and the multiplication rule $\mathbf{2}\otimes\mathbf{2}=\mathbf{3}\oplus\mathbf{1}$, thus allowing to reproduce the successful $U(2)$ textures \cite{Aranda:2000tm}. It has the following tensor product rules \cite{Ishimori:2010au}:
\begin{eqnarray}
\2tvec{x_1}{x_2}_{\bf 2(2')}\otimes\2tvec{y_1}{y_2}_{\bf 2(2'')}
=
\left(\frac{x_1y_2-x_2y_1}{\sqrt{2}}\right)_{\bf 1}
\oplus\3tvec{\frac{i}{\sqrt{2}}p_1p_2\bar p(x_1y_2+x_2y_1)}{p_2\bar p^2x_1y_1}{x_2y_2}_{\bf3}.
\end{eqnarray}
\begin{eqnarray}
\2tvec{x_1}{x_2}_{\bf2'(2)}\otimes\2tvec{y_1}{y_2}_{\bf2'(2'')}
&=\left(\frac{x_1y_2-x_2y_1}{\sqrt{2}}\right)_{\bf1''}
\oplus
\3tvec{p_1\bar p^2x_1y_1}{x_2y_2}{\frac{i}{\sqrt{2}}\bar p\bar
  p_2(x_1y_2+x_2y_1)}_{\bf3}, 
\\
\2tvec{x_1}{x_2}_{\bf2''(2)}\otimes\2tvec{y_1}{y_2}_{\bf2''(2')}
&=\left(\frac{x_1y_2-x_2y_1}{\sqrt{2}}\right)_{\bf1'}
\oplus
\3tvec{x_2y_2}{\frac{i}{\sqrt{2}}\bar p\bar p_1(x_1y_2+x_2y_1)}{\bar p^2\bar p_1\bar p_2x_1y_1}_{\bf3}.
\end{eqnarray}
\begin{eqnarray}
\bf2\times 2'=2''\times 2'', \qquad 2\times2''=2'\times 2', \qquad
2'\times 2''=2\times 2,
\end{eqnarray}
\begin{eqnarray}
\3tvec{x_1}{x_2}{x_3}_{\bf3}\otimes\3tvec{y_1}{y_2}{y_3}_{\bf3}
&=&
[x_1y_1+p^2_1p_2(x_2y_3+x_3y_2)]_{\bf1}\nonumber
\\&\oplus&
[x_3y_3+\bar p_1\bar p^2_2(x_1y_2+x_2y_1)]_{\bf1'}
\oplus
[(x_2y_2+\bar p_1p_2(x_1y_3+x_3y_1)]_{\bf1''}\nonumber
\\&\oplus&
\3tvec{2x_1y_1-p^2_1p_2(x_2y_3+x_3y_3)}
{2p_1p^2_2x_3y_3-x_1y_2-x_2y_1}{2p_1\bar p_2x_2y_2-x_1y_3-x_3y_1}_{\bf3_1}
\nonumber
\\&\oplus&
\3tvec{x_2y_3-x_3y_2}{\bar p^2_1\bar p_2(x_1y_2-x_2y_1)}
{\bar p^2_1\bar p_2(x_3y_1-x_1y_3)}_{\bf3_2},
\end{eqnarray}
where $p_1= e^{i\phi_1}$ and $p_2= e^{i\phi_2}$.
 
\section{Scalar potential for one of the $T'$ scalar triplets}
\label{scalarpotentialTprime}
The scalar potential for the $T'$ scalar triplet $\rho$ is given by:
\begin{align}\label{eqn:scalpott}
   V =& -\mu^{2}_{\rho}(\rho \rho^{*})_{\mathbf{1}} + \kappa_{1}(\rho \rho^{*})_{\mathbf{1'}}(\rho \rho^{*})_{\mathbf{1''}} + \kappa_{2}(\rho \rho^{*})_{\mathbf{1''}}(\rho \rho^{*})_{\mathbf{1'}} \nonumber \\ 
   &+ \kappa_{3}(\rho \rho^{*})_{\mathbf{3}_{1}}(\rho \rho^{*})_{\mathbf{3}_{1}} + \kappa_{4}(\rho \rho^{*})_{\mathbf{3}_{2}}(\rho \rho^{*})_{\mathbf{3}_{2}} + \kappa_{5}(\rho \rho^{*})_{\mathbf{3}_{1}}(\rho \rho^{*})_{\mathbf{3}_{2}} + h.c.
\end{align}

This scalar potential has six free parameters: one bilinear and five quartic couplings. The $\mu_{\rho}$ parameter can be written as a function of the other five parameters by the scalar potential minimization condition:
\begin{align}
    \frac{\partial \langle V(\rho) \rangle}{\partial v_{\rho}} =&\ 8 \left(\cos (\gamma )-e^{-i \left(2 \phi _1+\phi _2\right)} \sin (\gamma )\right)^{2} \left(\cos (\gamma )-e^{i \left(2 \phi _1+\phi _2\right)} \sin
   (\gamma )\right){}^2 \kappa _1 v _{\rho }^3 \nonumber \\ & +8 \left(\cos (\gamma )-e^{-i \left(2 \phi _1+\phi _2\right)} \sin (\gamma )\right){}^2 \left(\cos
   (\gamma )-e^{i \left(2 \phi _1+\phi _2\right)} \sin (\gamma )\right){}^2 \kappa _2 v _{\rho }^3 \nonumber \\ &+32 \left(\cos (\gamma )-e^{-i \left(2 \phi
   _1+\phi _2\right)} \sin (\gamma )\right){}^2 \left(\cos (\gamma )-e^{i \left(2 \phi _1+\phi _2\right)} \sin (\gamma )\right){}^2 \kappa _3 v
   _{\rho }^3 \nonumber \\ &+2 \left(\cos (\gamma )-e^{-i \left(2 \phi _1+\phi _2\right)} \sin (\gamma )\right) \left(\cos (\gamma )-e^{i \left(2 \phi _1+\phi
   _2\right)} \sin (\gamma )\right) \nonumber \\ &\times \kappa _5 \left(2 e^{i \alpha } \left(\cos (\gamma )-e^{-i \left(2 \phi _1+\phi _2\right)} \sin (\gamma )\right)
   v _{\rho } -2 e^{-i \alpha } \left(\cos (\gamma )-e^{i \left(2 \phi _1+\phi _2\right)} \sin (\gamma )\right) v _{\rho }\right) v _{\rho }^2 \nonumber \\ &+2
   \left(\cos (\gamma )-e^{-i \left(2 \phi _1+\phi _2\right)} \sin (\gamma )\right) \left(\cos (\gamma )-e^{i \left(2 \phi _1+\phi _2\right)} \sin
   (\gamma )\right) \nonumber \\ & \times \kappa _5 \left(2 e^{-i \alpha } \left(\cos (\gamma )-e^{i \left(2 \phi _1+\phi _2\right)} \sin (\gamma )\right) v _{\rho }-2
   e^{i \alpha } \left(\cos (\gamma )-e^{-i \left(2 \phi _1+\phi _2\right)} \sin (\gamma )\right) v _{\rho }\right) v _{\rho }^2 \nonumber \\ & -4 \left(\cos
   (\gamma )-e^{-i \left(2 \phi _1+\phi _2\right)} \sin (\gamma )\right) \left(\cos (\gamma )-e^{i \left(2 \phi _1+\phi _2\right)} \sin (\gamma
   )\right) \mu _{\rho }^2 v _{\rho } \nonumber \\ & +4 \left(\cos (\gamma )-e^{-i \left(2 \phi _1+\phi _2\right)} \sin (\gamma )\right) \left(\cos (\gamma )-e^{i
   \left(2 \phi _1+\phi _2\right)} \sin (\gamma )\right) \nonumber \\ &\times  \kappa _5 \left(e^{i \alpha } \left(\cos (\gamma )-e^{-i \left(2 \phi _1+\phi _2\right)}
   \sin (\gamma )\right) v _{\rho }^2-e^{-i \alpha } \left(\cos (\gamma )-e^{i \left(2 \phi _1+\phi _2\right)} \sin (\gamma )\right) v _{\rho
   }^2\right) v _{\rho }\nonumber \\ & +4 \left(\cos (\gamma )-e^{-i \left(2 \phi _1+\phi _2\right)} \sin (\gamma )\right) \left(\cos (\gamma )-e^{i \left(2 \phi
   _1+\phi _2\right)} \sin (\gamma )\right) \nonumber \\ &\times \kappa _5 \left(e^{-i \alpha } \left(\cos (\gamma )-e^{i \left(2 \phi _1+\phi _2\right)} \sin (\gamma
   )\right) v _{\rho }^2-e^{i \alpha } \left(\cos (\gamma )-e^{-i \left(2 \phi _1+\phi _2\right)} \sin (\gamma )\right) v _{\rho }^2\right) v
   _{\rho } \nonumber \\ & +2 \kappa _4 \left(2 e^{i \alpha } \left(\cos (\gamma )-e^{-i \left(2 \phi _1+\phi _2\right)} \sin (\gamma )\right) v _{\rho }-2 e^{-i
   \alpha } \left(\cos (\gamma )-e^{i \left(2 \phi _1+\phi _2\right)} \sin (\gamma )\right) v _{\rho }\right) \nonumber \\ & \times\left(e^{i \alpha } \left(\cos
   (\gamma )-e^{-i \left(2 \phi _1+\phi _2\right)} \sin (\gamma )\right) v _{\rho }^2-e^{-i \alpha } \left(\cos (\gamma )-e^{i \left(2 \phi _1+\phi
   _2\right)} \sin (\gamma )\right) v _{\rho }^2\right) \nonumber \\ &+2 \kappa _4 \left(2 e^{-i \alpha } \left(\cos (\gamma )-e^{i \left(2 \phi _1+\phi
   _2\right)} \sin (\gamma )\right) v _{\rho }-2 e^{i \alpha } \left(\cos (\gamma )-e^{-i \left(2 \phi _1+\phi _2\right)} \sin (\gamma )\right) v
   _{\rho }\right) \nonumber \\ &\times \left(e^{-i \alpha } \left(\cos (\gamma )-e^{i \left(2 \phi _1+\phi _2\right)} \sin (\gamma )\right) v _{\rho }^2-e^{i \alpha }
   \left(\cos (\gamma )-e^{-i \left(2 \phi _1+\phi _2\right)} \sin (\gamma )\right) v _{\rho }^2\right) \nonumber \\ &=0
\end{align}

Here for the sake of simplicity we consider vanishing phases in the multiplications rules for the tensor product of the scalar triplets of $T'$. Then, the scalar potential minimization condition yields the following relation: 
\begin{equation}
    \mu_{\rho}^{2} = v _{\rho }^2 \left(\frac{16 \kappa _4 \left(\sin (\alpha ) \cos (\gamma )-\sin (\gamma ) \sin \left(\alpha -2 \phi _1-\phi
   _2\right)\right){}^2}{\sin \left(2 \gamma -2 \phi _1-\phi _2\right)+\sin \left(2 \left(\gamma +\phi _1\right)+\phi _2\right)-2}-2 \left(\kappa
   _1+\kappa _2+4 \kappa _3\right) \left(\sin (2 \gamma ) \cos \left(2 \phi _1+\phi _2\right)-1\right)\right)
\end{equation}

This result indicates that the VEV pattern of the $T'$ triplet $\rho$ in \eqref{eqn:vevt} is consistent with a global minimum of the scalar potential \eqref{eqn:scalpott} of this model for a large region of parameter space. Following the same procedure previously described, one can also show that the VEV patterns of the $T'$ triplets $\phi$, $\zeta$ and $\xi$ in \eqref{eqn:vevt} are also consistent with the scalar potential minimization equations.

\end{document}